\documentclass[%
reprint,
superscriptaddress,
 amsmath,amssymb,
 aps,
pra,
]{revtex4-1}
\usepackage{graphicx}
\graphicspath{{./Figures/}}
\usepackage{graphicx}
\usepackage{dcolumn}
\usepackage{amsmath}
\usepackage{bm}
\usepackage{booktabs} 
\usepackage{multirow}
\usepackage{physics}
\usepackage{caption}
\usepackage{subcaption}
\usepackage{hyperref}
\usepackage{siunitx} 

\usepackage{braket}
\usepackage{color}
\usepackage{subcaption}
\usepackage{soul}

\begin{document}
\bibliographystyle{apalike}
\graphicspath{{./Figures/}}
\preprint{APS/123-QED}

\title{ Charge qubits based on ultra-thin topological insulator films}
\author{K. X. Zhang}
\email{echokxz@zhejianglab.com}
\affiliation{Research Center for Quantum Sensing, Intelligent Perception Research Institute, Zhejiang Lab, Hangzhou, 311121, China}

\affiliation{Cavendish Laboratory, Department of Physics, University of Cambridge, Cambridge CB3 0HE, United Kingdom}%
\author{Hugo V. Lepage}%
\affiliation{Cavendish Laboratory, Department of Physics, University of Cambridge, Cambridge CB3 0HE, United Kingdom}%
\author{Ying Dong}
\affiliation{Research Center for Quantum Sensing, Intelligent Perception Research Institute, Zhejiang Lab, Hangzhou, 311121, China}
\author{C. H. W. Barnes}
\email{chwb101@cam.ac.uk}
\affiliation{Cavendish Laboratory, Department of Physics, University of Cambridge, Cambridge CB3 0HE, United Kingdom}%

\begin{abstract}
We study how to use the surface states in a Bi\textsubscript{2}Se\textsubscript{3} topological insulator ultra-thin film that are affected by finite size effects for the purpose of quantum computing. We demonstrate that: (i) surface states under the finite size effect can effectively form a two-level system where their energy levels lie in between the bulk energy gap and a logic qubit can be constructed, (ii)  the qubit can be initialized and manipulated using electric pulses of simple forms, (iii) two-qubit entanglement is achieved through a $\sqrt{\text{SWAP}}$ operation when the two qubits are in a parallel setup, and (iv) alternatively, a Floquet state can be exploited to construct a qubit and two Floquet qubits can be entangled through a Controlled-NOT operation. The Floquet qubit offers robustness to background noise since there is always an oscillating electric field applied, and the single qubit operations are controlled by amplitude modulation of the oscillating field, which is convenient experimentally.

\end{abstract}

\maketitle

\section{Introduction}
Topological insulators (TIs), as a recently discovered material, have created a research surge, owing to their intriguing surface states \cite{Ando2013a,Hasan2010,Hasan2011}. As a result of the preservation of time-reversal symmetry and the strong spin-orbit coupling effect in the material, the Hilbert space of a TI poses a non-trivial topology and surface states appear when interfacing with a trivial insulator (eg. air). These surface states are unique, in a sense that their existence is protected by the time-reversal symmetry and thus they are robust against non-magnetic local perturbations. TI surface states show interesting properties such as anti-weak localization \cite{Lu2014} and spin-momentum locking \cite{Ando2013a}. The high mobility and low power dissipation make TIs ideal for transmitting information. All of these benefits make TIs promising for applications in fault-tolerant quantum computing and spintronics \cite{Hasan2010,Pesin2012,He2019}, and there has been continuous advances in this area \cite{Hasan2010,Cho2012,Herath_2014,PhysRevB.98.165430,doi:10.1063/1.4978632,PhysRevB.84.233101,PhysRevLett.111.106802,castroenriquez2020optical,PhysRevB.99.245407}.  
In order to exploit the uses of a TI, it is essential to functionalize the material. The fabrication of TI nanodevices is an active area of research\cite{Xiu_2013,Liu_2020,Li2012,HU2014}. Among the nanodevices, TI quantum dots or ultra-thin TI films are interesting areas of research due to their small size, low dimension, and possibility to host a qubit. Recently, a TI quantum dot was successfully fabricated using $\text{Bi}_{2}\text{Se}_{3}$ with tunable barriers controlled by gate voltages \cite{Cho2012}. The energy spectrum of a TI quantum dot and optical transitions between the bands have been studied theoretically \cite{Herath_2014}. The decoherence mechanism of a TI is under investigation \cite{PhysRevB.99.245407}. A magnetically-defined TI qubit has been studied theoretically \cite{PhysRevLett.111.106802}. When the thickness of a TI device is reduced to a few nanometers, the finite size effect of a nanodevice causes the hybridization of the surface states located on opposite surfaces and a hybridized gap opens at its surface. This is an essential property to the application of TI devices since it allows the formation of a two-level system which is energetically separate from other bulk states. Ultra-thin TI films show interesting properties such as the oscillation between a quantum spin Hall phase \cite{PhysRevB.81.041307} and an ordinary insulator phase depending on its thickness, and the spin can be selectively injected by circular polarized lights \cite{PhysRevB.81.115407}. 

In recent years, the fast development of laser and ultra-fast spectroscopy techniques have enabled scientists to gain more control of a quantum system that is out of equilibrium. Floquet engineering - the subject of controlling a quantum system with time-periodic driving fields, has benefited from this technical development. It has provided a useful tool to study non-equilibrium systems \cite{Oka2019}, and interests in this field is growing fast \cite{Oka2019, Kolodrubetz2018, Oka2009, Bilitewski2015}. Floquet-Bloch states on the surface of a topological insulator were observed in 2013 \cite{Wang2013}. 

In this work, we explore the possibilities of making a TI qubit in an ultra-thin TI film and examine whether the system fulfils DiVincenzo's criteria, which are necessary for making a TI quantum computer. Moreover, we explore a Floquet-engineered qubit in a TI system, which benefits from oscillating fields to gain extra robustness against background charge fluctuations \cite{PhysRevA.100.012341}. In section \ref{sec:charq}, we investigate the initialization and controllability of a single qubit and find that the qubit can be rotated to an arbitrary position on the Bloch sphere by a unitary quantum operation defined by a sequence of carefully designed square pulses. In section \ref{sec:flqu}, we investigate a Floquet TI qubit that accomplishes the same task with cosine pulses. The Floquet qubit can also be rotated to an arbitrary position on the Bloch sphere, independently of its initial position. This rotation can be achieved by tuning the amplitude of the driving field, so-called amplitude modulation. 
In section \ref{sec:twoq}, we study the entanglement generation of two qubits in neighbouring TI thin films for the charge qubit and the Floquet qubit cases separately. 
We find that the two qubits can be entangled through $\sqrt{\text{SWAP}}$ gates by individually controlling their energy levels using simple electric pulses. We find that the bands decouple and there is no SWAP gates if the inner dot tunneling is too strong comparing to the electrostatic coupling strength between the qubits \cite{ lepage2020entanglement}. On the other hand, two Floquet qubits can be entangled through a Controlled-NOT (CNOT) gate by aligning the two qubits at a short distance with simple electric pulses applied on them individually.

\section{Hamiltonian And Numerical Methods}
A topological insulator thin film with an external electric field is simulated using Liu's model Hamiltonian \cite{PhysRevB.82.045122} with an extra term that represents a homogeneous electric field perpendicular to the thin film.  The Hamiltonian is:
\begin{equation}\label{eq:cH}
  H = C(\mathbf{k})I_4+M(\mathbf{k})\Gamma_5+B_0\Gamma_4k_z+A_0(\Gamma_1k_y-\Gamma_2k_x)-e\mathbf{E}\hat{\mathbf{z}}I_4
\end{equation}.

$I_4$ is a $4\cross4$ identity matrix and:
\begin{align*}
&\Gamma_5 =\begin{pmatrix}
1 & 0 & 0 & 0\\
0 & -1 & 0 & 0\\
0& 0 & 1 & 0\\
0 & 0 & 0 & -1 
\end{pmatrix} ,\
\Gamma_4 =\begin{pmatrix}
0 & -i & 0 & 0 \\
i & 0 & 0 & 0 \\
0 & 0 & 0 & -i \\
0 & 0 & i & 0 
\end{pmatrix} ,\\
&\Gamma_1 =\begin{pmatrix}
0 & 0 & 0 & -i\\
0 & 0 & -i & 0\\
0 & i & 0 & 0\\
i & 0 & 0 & 0\\  
\end{pmatrix} ,\
\Gamma_2 =\begin{pmatrix}
0 & 0 & 0 & 1\\
0 & 0 & 1 & 0\\
0 & 1 & 0 & 0\\
1 & 0 & 0 & 0\\  
\end{pmatrix} ,
\end{align*}
\begin{align*}
&C(\mathbf{k})=C_0+C_1k_z^2+C_2|k_{||}|^2 ,\\ &M(\mathbf{k})=M_0+M_1k_z^2+M_2|k_{||}|^2 ,\\
&k_{||} = k_x-i k_y.
\end{align*}

The last term in Eq. \ref{eq:cH}, $-e\mathbf{E}\hat{\mathbf{z}}$, stands for the homogeneous electric field that is perpendicular to the thin film. The parameters $C_0$, $C_1$, $C_2$, $M_0$, $M_1$, $M_2$, $B_0$, $A_0$ are determined by material properties and are obtained by first-principles calculations. 
In this article, we use the values from Liu's paper \cite{PhysRevB.82.045122}. The time evolution of a qubit is studied using the Staggered Leapfrog method \cite{Visscher1991, arvidsson2017protocol, lepage2020fermionic, takada2019sound}.  

\subsubsection{The Floquet-engineered charge qubit} 
The Hamiltonian with a time-periodic electric field $\mathbf{E}=E_{0}\cos(\omega t)\hat{\mathbf{z}}$ is diagonalizable using the Floquet method \cite{Shirley1965}. Floquet theory is suitable for studying time-periodic systems.  

The general periodic time-dependent Schrödinger equation with a period $T$ is written as:
\begin{equation}\label{eq:floq}
  i\hbar\frac{\partial \Phi(t)}{\partial t}= H\Phi(t),
\end{equation}

with $H(t+T) = H(t)$. According to Floquet theory, there exist time-dependent solutions to Eq. \ref{eq:floq} with the form:
\begin{equation}\label{eq:floq2}
	  \Phi_{\alpha}(t)= e^{-i\epsilon_\alpha t/\hbar}u_{\alpha}(t).
\end{equation}

$\epsilon_\alpha $ is the quasi energy of state $\alpha$, called the Floquet energy and $u_{\alpha}(t)$ is a periodic function with $u_{\alpha}(t+T) = u_{\alpha}(t)$. 
The solution is similar to a Bloch solution in a spatially periodic system, with a phase term of $e^{-i\epsilon_\alpha t/\hbar}$ and a periodic function $u_{\alpha}(t)$. Substituting Eq. \ref{eq:floq2} into Eq. \ref{eq:floq}, we arrive at:

\begin{equation} \label{eq:floq3}
( H-i\hbar\frac{\partial}{\partial t})u_{\alpha}(t)=\epsilon_{\alpha} u_{\alpha}(t).
\end{equation}

The LHS, $H-i\hbar\frac{\partial}{\partial t}$, is called the Floquet Hamiltonian. Because the Floquet state $u_{\alpha}$ is periodic, we can rewrite it as a Fourier series $u_{\alpha}(t) = \sum_{n=-\infty}^{n=\infty} a_n e^{in\omega t}$. $a_n$ are the Fourier coefficients and $\omega$ is the frequency of the driving system. Therefore, the explicit time dependence of Eq. \ref{eq:floq3} is replaced by the Fourier representation and this Floquet Hamiltonian can be viewed as a time-independent Hamiltonian with quasi energies $\epsilon_{\alpha}+ n\hbar\omega$.

\section{Results and Discussion}

\subsection{Charge qubit}\label{sec:charq}
\begin{figure}[htbp]
    \centering
   \includegraphics[width=0.5\textwidth]{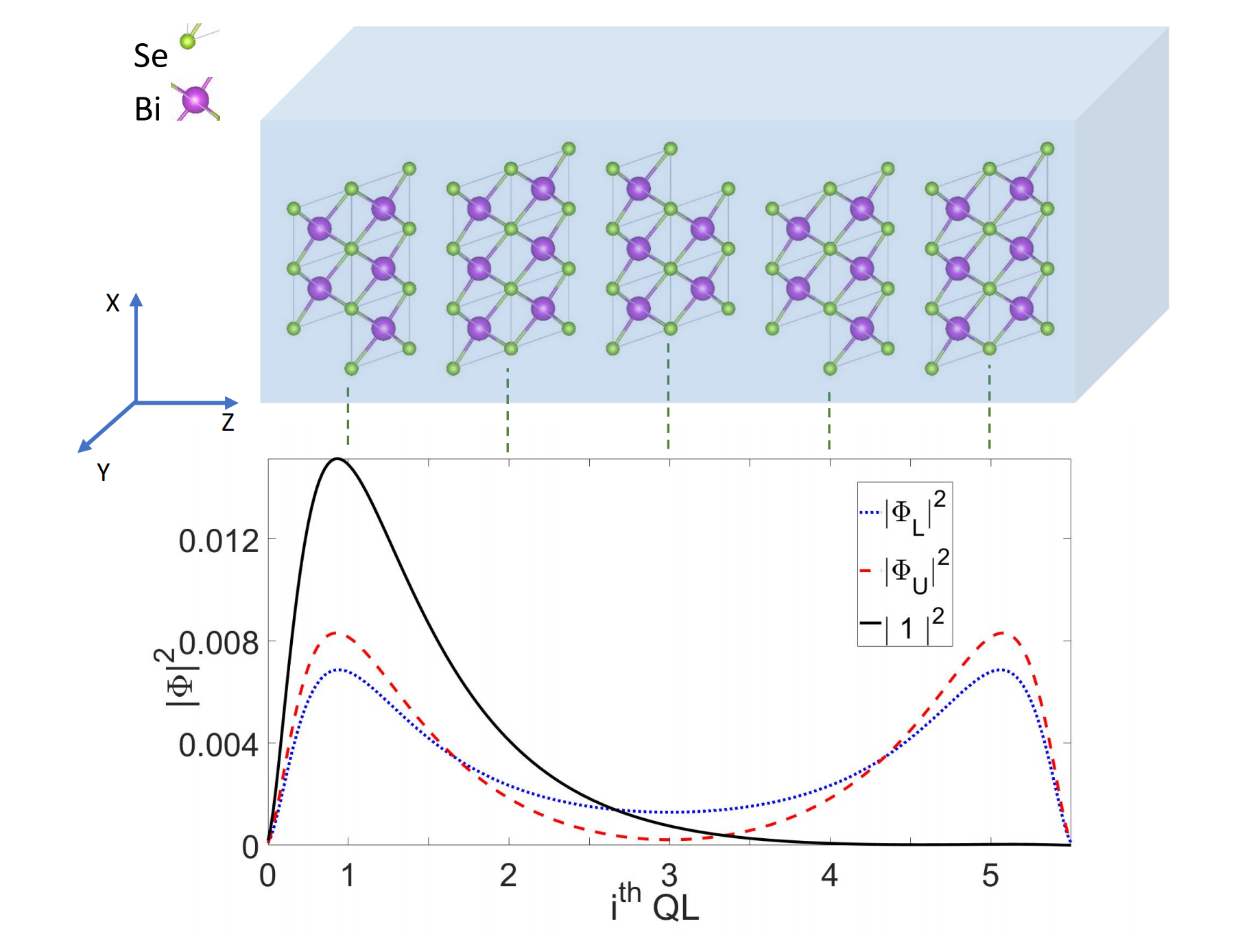}
   \caption{Top: graphical representation of five Bi$_2$Se$_3$ QLs. The wave densities of the hybridized surface states and the combined qubit state $\ket{1}$ in a 5-QL TI slab (finite in the $z$ direction). The red dashed curve is anti-bonding state $\Phi_U$ located at the upper energy level of the gapped Dirac Cone, while the blue dotted curve is the bonding state $\Phi_L$ located at the lower energy  level of the gapped Dirac Cone. The black solid curve is the combined qubit state.}
    \label{fig:cq_1}
\end{figure}
A 3D topological insulator has a Dirac-cone band structure near its surface. The eigenstates on the Dirac cone are called surface states. Their spins are aligned with the surface and perpendicular to the direction of motion of the electron. At the $\Gamma$ point ($k_{x}=k_{y}=0$), there are four surface states with spins perpendicular to the surface ($s_z=\pm\frac{1}{2}$). When the thickness of the TI slab is reduced, the surface states at the $\Gamma$ point overlap and tunnelling occurs between each surface. The hybridization of the surface states open up a gap in the Dirac cone, which is observed to increase with an oscillatory pattern when the thickness of a TI ultra-thin film is reduced in \cite{Linder2009,PhysRevB.81.041307, PhysRevB.81.115407}. In this article, we use a five-quintuple-layer (QL) TI thin film with a hybridized gap of $\SI{0.1}{eV}$ throughout the whole article. These surface states are like the bonding and anti-bonding states found in a pseudo molecule (Fig. \ref{fig:cq_1}). We define the surface states at the $\Gamma$ point on the upper part of the gapped Dirac cone as $\ket{\Phi_{U}}$ and those at the $\Gamma$ point on the lower part of the gapped Dirac cone as $\ket{\Phi_{L}}$. The spin-degenerate pair of $\ket{\Phi_{U}}$ (or $\ket{\Phi_{L}}$) cannot be scattered into each other in the presence of elastic scattering since they have opposite spin components, unless in the case of a magnetic impurity.  Therefore, the robustness of these state are guaranteed against elastic scattering, which is the major scattering mechanism in a realistic device at low temperature. Note that the electric pulses only address the states with the same spin. By combining a pair of surface states at the $\Gamma$ point using an electric field or a surface gate, we obtain a pair of states located on the top/bottom surfaces of the TI ultra-thin film (Fig. \ref{fig:cq_1}). We define these as our charge qubit states. The logical state $\ket{0}$ is the state located on the top surface and $\ket{1}$ is located on the bottom surface (the solid black line in Fig. \ref{fig:cq_1}). 

 \begin{align} 
\ket{0} = \frac{1}{\sqrt{2}}(\ket{\Phi_L}+\ket{\Phi_U}),\label{eq:char0}\\
\ket{1}=\frac{1}{\sqrt{2}}(\ket{\Phi_L}-\ket{\Phi_U}).\label{eq:char1}
\end{align}

All single-qubit calculations are modelled using the Hamiltonian in Eq. \ref{eq:cH} with the bands near the Fermi level and we can see the qubit is always on the Bloch sphere during the initialization and the operation process. Therefore, we can confirm this is a pure two-level system under the proposed manipulation. This two-level system can be considered like a qubit in a quantum dot.

\begin{figure}[htbp]
    \includegraphics[width=0.5\textwidth]{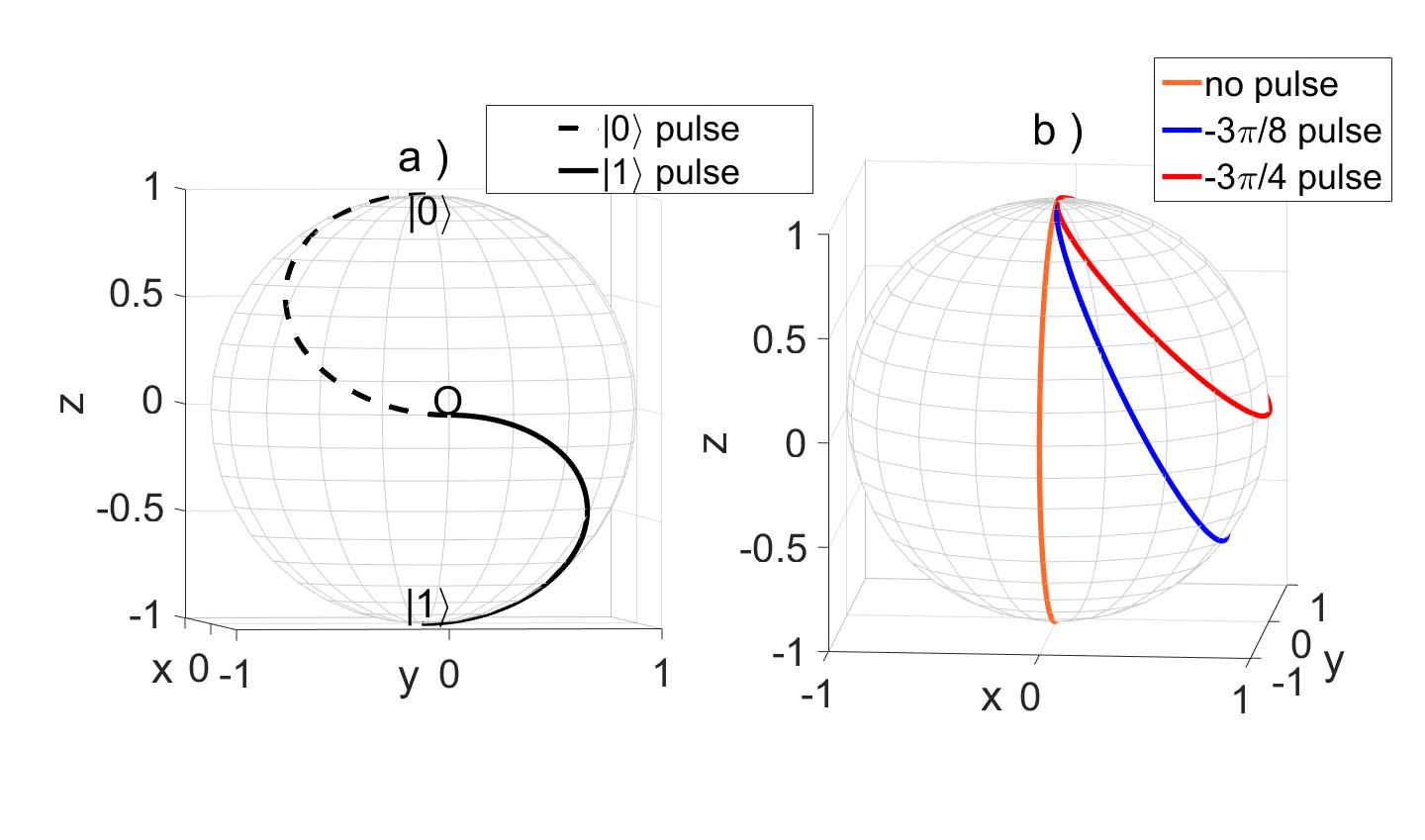}
   \caption{The Bloch sphere representation of basic rotation. \textbf{a)} the initialization from $\Phi_0$ to $\ket{0}$ and $\ket{1}$. The initialization to $\ket{0}$ is achieved with a static electric field applied in the $-z$ direction and the initialization to $\ket{1}$ is achieved with a static electric field applied in the $+z$ direction. \textbf{b)} the relation between the field amplitude and the axes of rotations. The axis of rotation tilts from the $x$ axis as the electric field increases; it always located in the $XZ$ plane.}
    \label{fig:cq_2}
\end{figure}

\subsubsection{Initialization}
The charge qubit can be initialized using an electric field pulse. A static electric field, generated by a gate voltage, perpendicular to the TI thin film (i.e in the $z$ direction) can be used to initialize the qubit to the state $\ket{0}$ or $\ket{1}$ depending on the direction of the field (see a in Fig \ref{fig:cq_2}). Alternatively, an oscillating electric field can be used to initialize the qubit. The initialization using an oscillating electric field will be discussed in the next section of Floquet-engineered charge qubits. We found that a static electric field will generate a single qubit rotation about an axis $\hat{n}$ in the $XZ$ plane. The amplitude of the electric field will determine the orientation of $\hat{n}$ (see b in Fig. \ref{fig:cq_2}). When no field is applied, the rotation follows a circular path about the $-x$ axis. When an electric field is applied in the $+z$ direction, $\theta_{\hat{n}} \in (0,\frac{\pi}{2})$ and $\phi_{\hat{n}} = \pi$ in spherical coordinates. When an electric field is applied in the $-z$ direction, $\theta_{\hat{n}} \in (\frac{\pi}{2},\pi)$ and $\phi_{\hat{n}} = \pi$. Therefore, changing the electric field amplitude only changes $\theta_{\hat{n}}$. For the convenience of description, a rotation with respect to the axis $\theta_{\hat{n}} \in (-\pi, 0)$ always refers to one with $\phi_{\hat{n}} = \pi$ and $\theta_{\hat{n}}\in (0,{\pi})$ in the followings unless stated otherwise. In order to initialize and control the rotation of a single charge qubit, the electric pulse of axes with $\theta_{\hat{n}}=-\frac{\pi}{4}$ and $\theta_{\hat{n}}=-\frac{3\pi}{4}$ are used in this paper. The pulses are found by sweeping over a range of electric field amplitudes. Given the fact that $\theta_{\hat{n}} = -\frac{\pi}{2}$ when there is no pulse, the axes tilt from the $-x$ axis by the same amount in the opposite direction when the electric pulse is flipped. An electric pulse with an amplitude in the perturbation range can be used to produce rotations about two orthogonal axes by flipping its direction, which is sufficient for arbitrary rotations. We found that with a proper electric field, we can produce a rotation about $\theta_{\hat{n}}=-\frac{\pi}{4}$ ( $\theta_{\hat{n}}=-\frac{3\pi}{4}$ ) and this can drive the qubit to the state $\ket{1}(\ket{0})$ at a half period of the rotation (see a in Fig. \ref{fig:cq_2}). In real situations, the rotational axes that an electric field produces depends on the material parameters, the thickness of the TI thin film and the amplitude of the field.   

\begin{figure*}[htbp]
    \includegraphics[width=1\textwidth]{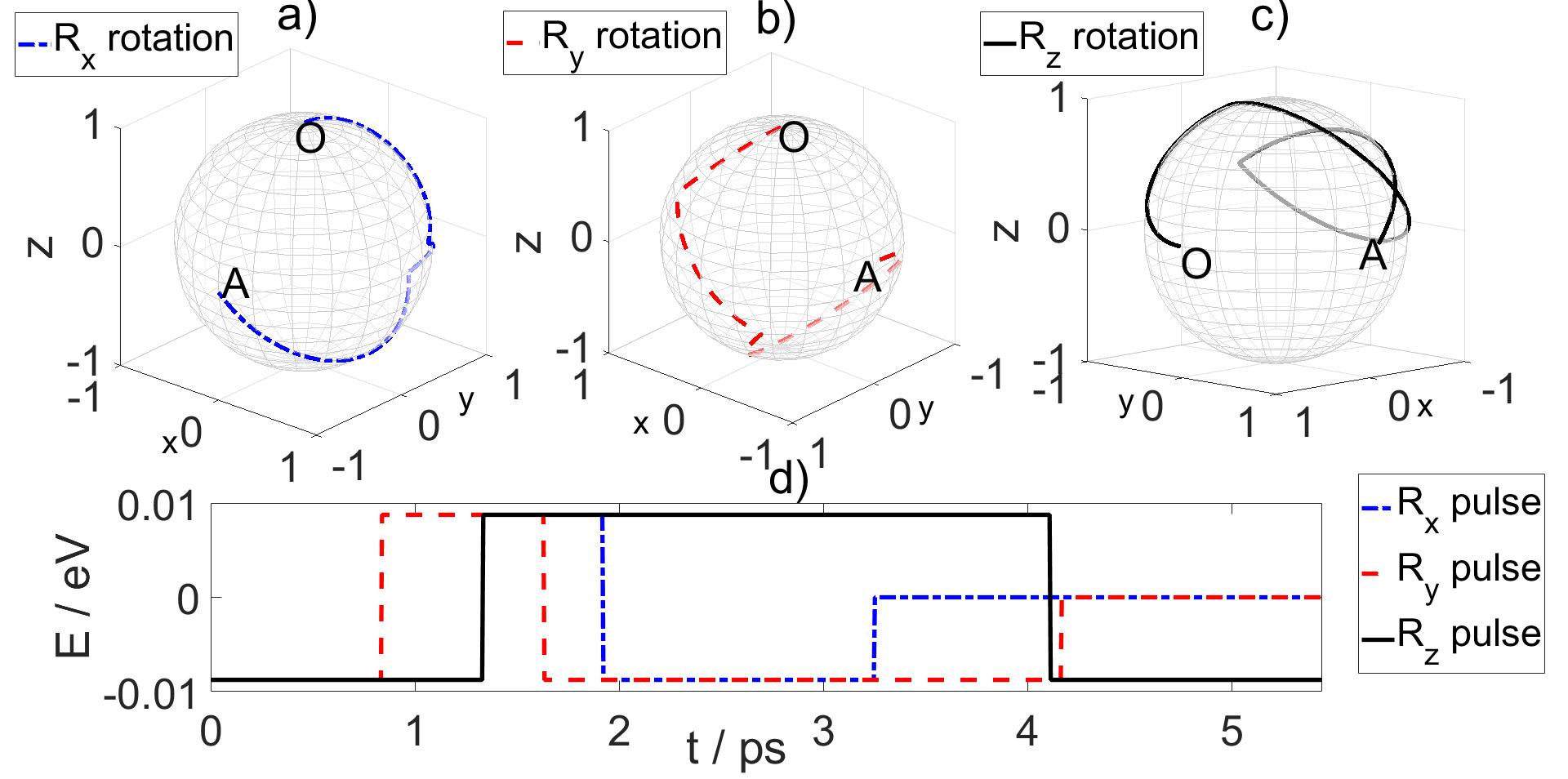}
	\caption{The Bloch sphere representation of  $R_x$, $R_y$ and $R_z$ rotations of angle $\SI{95}{^\circ}$ from $O$ to $A$ and the corresponding pulses. \textbf{a)} the path of the $R_x$ rotation, \textbf{b)} the path of the $R_y$ rotation, and \textbf{c)} the path of the $R_z$ rotation on the Bloch sphere. \textbf{d)} the pulses used to generate each rotation. The pulse times are calculated in picoseconds. The $R_z$ rotation is longer to achieve than the $R_x$ and $R_y$ rotations in the case of $\gamma = \SI{95}{^\circ}$ }
	\label{fig:cq_3}
\end{figure*}
\subsubsection{Single-Qubit Control} 
A universal rotation on the Bloch sphere can be represented by a unitary quantum gate. 

In general, a rotation by an angle $\gamma$ about an axis $\hat{n}$ can be expressed as:
\begin{align}
R_{\hat{n}}(\gamma) &= exp(-i\frac{\gamma}{2}\hat{n}\cdot\vec{\sigma}),\\
&= \cos(\frac{\gamma}{2})I_2-i\sin(\frac{\gamma}{2})\hat{n}\cdot\vec{\sigma}.
 \label{eq:geR}
\end{align}
$\vec{\sigma}$ is the vector of Pauli Matrices ($\sigma_x$, $\sigma_y$, $\sigma_z$) and $I_2$ is the $2\cross2$ identity matrix. $\hat{n} = (\cos\phi\sin\theta, \sin\phi\sin\theta,\cos\theta)$

According to Euler's theorem, we know that any arbitrary rotations can be achieved using three elementary rotations (e.g. about the $x$, $y$, $z$ axes). 
We extend the idea and define the rotation axes $\hat{n}_{-\frac{\pi}{4}}$ as $X'$ and $\hat{n}_{-\frac{3\pi}{4}}$ as $Z'$, where $\hat{n}_{-\frac{\pi}{4}}$ is the rotation axis at $\theta_{\hat{n}}=-\frac{\pi}{4}$ and $\hat{n}_{-\frac{3\pi}{4}}$ is the rotation axis at $\theta_{\hat{n}}=-\frac{3\pi}{4}$ respectively \cite{lasek2023pulse}. We choose the Euler angles representation for $R_{\hat{n}(\gamma)}$ as:
\begin{equation}
    R_{\hat{n}}(\gamma) = Z'(\beta_1)X'(\beta_2)Z'(\beta_3).
    \label{eq:euR}
\end{equation}
Substituting the RHS of Eq. \ref{eq:geR} with Eq. \ref{eq:euR}, we have:
\begin{equation}
  \cos(\frac{\gamma}{2})I_2-i\sin(\frac{\gamma}{2})\hat{n}\cdot\vec{\sigma}= Z'(\beta_1)X'(\beta_2)Z'(\beta_3).
\label{eq:R1}
\end{equation}
In this article, we aim to obtain the three elementary rotations: the $R_x$, $R_y$, and $R_z$ rotations about the $x$, $y$ and $z$ axes. From these, any single qubit quantum gates can be constructed. The $R_x$ rotation can be obtained for free (see Fig. \ref{fig:cq_2}), but here we show how to achieve it using pulse sequences. Expressing $\hat{x}$, $\hat{y}$, and $\hat{z}$ in a frame of $Z'$ and $X'$, we have:
\begin{align}
    \hat{x} &= (-\frac{1}{\sqrt{2}},0,-\frac{1}{\sqrt{2}}), \label{eq:tx1}\\
    \hat{y} &= (0,1,0),\\
    \hat{z} &= (\frac{1}{\sqrt{2}},0,-\frac{1}{\sqrt{2}}).
\end{align}
Then applying those separately in Eq. \ref{eq:R1}, we have for $R_x(\gamma)$:
\begin{align}
\beta_{1} &=\beta_{3} ,\\
\sin(\beta_{1}) &=\sqrt{\frac{1-\cos(\gamma)}{3+\cos(\gamma)}},\\
\cos(\beta_{1} ) &=-\frac{1}{\sqrt{2}}\frac{\sin(\gamma)}{\sin(\beta_{1})},\\
\cos(\beta_{2} ) &= \frac{1}{2}(\cos(\gamma)+1),\\
\sin(\beta_{2} ) &=\sin(\beta_{1})(1+\cos(\beta_{2})).
\end{align}

For $R_y(\gamma)$, we have:
\begin{align}
\beta_{1} &=\frac{\pi}{2},\\
\beta_{3} &=\frac{3\pi}{2},\\
\cos(\beta_{2}) &=\cos(\gamma),\\
\sin(\beta_{2} ) &=\sin(\gamma).
\end{align}

For $R_z(\gamma)$, we have:
\begin{align}
\beta_{1}&=\beta_{3},\\
\sin(\beta_{1}) &=\sqrt{\frac{1-\cos(\gamma)}{3+\cos(\gamma)}},\\
\cos(\beta_{1}) &=-\frac{\sqrt{2}\sin(\gamma)}{(3+\cos(\gamma))\sin(\beta_{1})},\\
\cos(\beta_{2}) &= \frac{1}{2}(\cos(\gamma)+1),\\
\sin(\beta_{2}) &=\frac{\sin(\gamma)}{\sqrt{2}\cos(\beta_{1})}.\label{eq:tz5}
\end{align}
With those equations, we achieve effective $R_x$, $R_y$, $R_z$ rotations of arbitrary angle $\gamma$ using a sequence of $R_{\hat{n}=-\frac{\pi}{4}}$ and $R_{\hat{n}=-\frac{3\pi}{4}}$ rotations. Because $\beta = \omega t$, and $\omega$ is the angular frequency that can be measured from experiments, we can obtain the time duration of the $R_{\hat{n}=-\frac{\pi}{4}}$ and $R_{\hat{n}=-\frac{3\pi}{4}}$ pulses to achieve a desired rotation of angle $\gamma$. 
A rotation about any axis on the Bloch sphere can be constructed by composing three rotations from these two axes $R_{\hat{n}=-\frac{\pi}{4}}$ and $R_{\hat{n}=-\frac{3\pi}{4}}$. We show some examples of  $R_x$, $R_y$, $R_z$ rotations in Fig. \ref{fig:cq_3}.

\subsubsection{Readout} 
The method of measuring this qubit after the operation is similar to measuring those in a double quantum dot - by connecting the qubit to a single-electron tunneling device, such as a single-electron transistor or a quantum point contact \cite{PhysRevLett.95.090502} . 

\subsection{Floquet-engineered charge qubit}\label{sec:flqu}
When a time-periodic field is applied to the quantum system, we are able to convert the time-dependent Hamiltonian to a time-independent Floquet Hamiltonian. If we tune the frequency of the electric field to match the Rabi frequency of the gapped Dirac cone, we are able to create a pair of Floquet states, which are combined states $a_1\ket{\Phi_L} \pm a_2\ket{\Phi_U}$, and $|a_1|^2+|a_2|^2 = 1$. The amplitude of the field determines the ratio $\mu = \frac{|a_1|}{|a_2|}$. The matching frequency $\omega = \Delta E/\hbar$, where $\Delta E \equiv E(\Phi_U)-E(\Phi_L)$ is the energy difference of the hybridized surface states in a bare TI system. The periodic electric field combines the electronic states with the same spin orientation and the spin of the resultant state is unchanged. 
\begin{figure*}[htbp]
    \centering
    \includegraphics[width=1\textwidth]{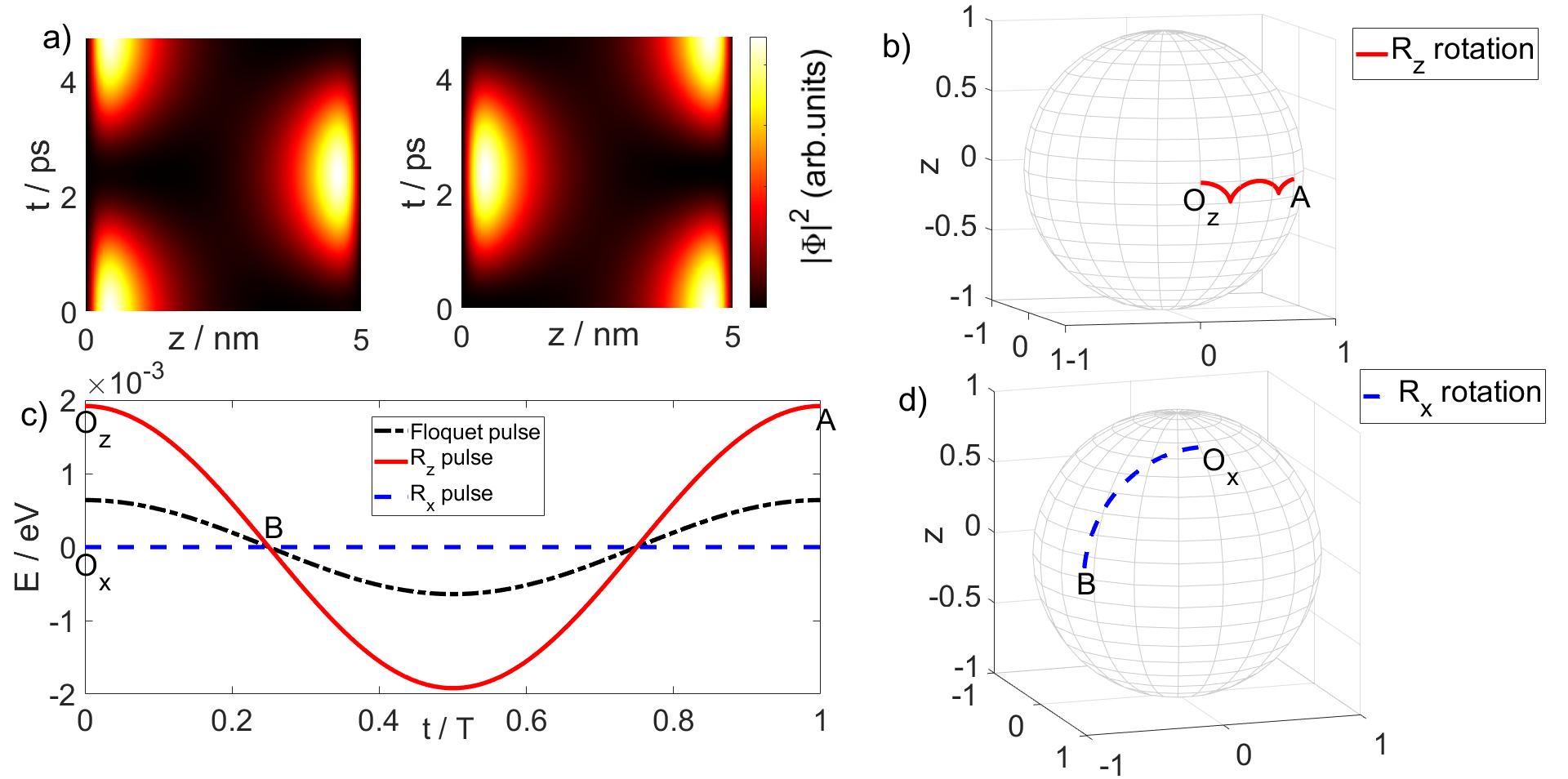}
   \caption{\textbf{a)} The electron density of $\ket{1(t)}$ (left) and $\ket{0(t)}$ (right) of a Floquet qubit vs time. \textbf{b)} The trajectory of a $R_z$ rotation from $O_z$ to $A$ in the Floquet frame. \textbf{c)} The pulses used to control the single qubit: initialization (black dash-dotted line), $R_x$ (blue dashed line), and $R_z$ (red solid line). \textbf{d)} The trajectory of a $R_x$ rotation from $O_x$ to $B$ in the static frame.}
    \label{fig:floq1}
\end{figure*}

\subsubsection{Initialization} 
A time-periodic electric field oscillating at a frequency at the Rabi frequency of the gap will excite an electron into the combined states $\ket{\Phi_{\text{FR}}}(\ket{\Phi_{\text{FL}}}) =a_1\ket{\Phi_L} +(-) a_2\ket{\Phi_U}$ at time $t=0$, where $\ket{\Phi_{\text{FL}}}$ is the Floquet state located in the $1^{st}$ quintuple layer (QL) and $\ket{\Phi_{\text{FR}}}$ is the Floquet state located in the $5^{th}$ QL. The electric field amplitude will determine $\mu = \frac{|a_1|}{|a_2|}$. Here, we use a pair of Floquet states with $\mu = 1$ as our Floquet-engineered charge qubit. This state would evolve as the charge qubit at $\ket{0}(\ket{1})$ in Eq. \ref{eq:char0} (\ref{eq:char1}). It is worth mentioning that the Floquet-engineered qubit exists in a TI with an oscillating field, therefore it is time dependent and periodic. The logic qubit states are $\ket{0(t)}$ and $\ket{1(t)}$ (Fig. \ref{fig:floq1} a). The Bloch sphere of a Floquet-engineered qubit rotates periodically. 

\subsubsection{Single Qubit Control}
Any single-qubit operation applied to a Floquet-engineered charge qubit is with respect to the time-varying Bloch sphere. In fact, the rotation of the Bloch sphere makes the single qubit rotations easier to obtain than the case of a charge qubit.  As mentioned previously, two orthogonal rotations on the Bloch sphere are sufficient for realizing universal single qubit gates.  We first look at the $R_x$ rotation. This rotation can be obtained by relabelling the time duration of a qubit (replace $t_1$ to $t=0$). In this way, the initial position (at $t=0$) of a qubit is replaced by its position at $t=t_1$ (Fig. \ref{fig:floq1} d)). $R_z$ is convenient to achieve using amplitude modulation techniques \cite{7435074}. We find that the trajectory of a rotation of the Floquet-engineered charge qubit overlaps with the trajectory of a $R_z$ rotation at $t = \frac{nT}{2}$, where $T$ is the period of the driving field and $n\in Z$ (Fig. \ref{fig:floq1} b). Therefore, one can simply tune the amplitude of the driving field to apply a $R_z$ rotation to the qubit (Fig. \ref{fig:floq1} c).

\subsubsection{Readout} 
The measurement is similar to the one for a charge qubit, which can be achieved by a single-electron transistor or a quantum point contact \cite{PhysRevLett.95.090502}. The only difference is that now we should be aware that the qubits are time-dependent, therefore, the initial time and the measuring time of the qubit should be recorded. If there are multiple Floquet qubits, the initial and measuring time should be recorded for each of them.

\subsection{The two-qubit gate}\label{sec:twoq}
A two-qubit gate can be realized by fabricating two TI thin films next to each other. In this paper, we apply a parallel setup as shown in Fig.\ref{fig:setup}. The vertical setup is not favourable for our TI system owing to a weak Coulomb force between the qubits preventing two-qubit gates. Two-qubit entanglement can be generated by applying electric pulses to qubits 1 and  2 separately. We label the two-electron charge bases of the system as ${\ket{LL}, \ket{LR}, \ket{RL}, \ket{RR}}$, where $\ket{ij} = \ket{i}\otimes\ket{j}$ and $\ket{L} = \ket{1}, \ket{R} = \ket{0}$. From here we will discuss the results of the charge qubit and the Floquet qubit in turn.

\subsubsection{The Hamiltonian}
The two-qubit Hamiltonian using the Pauli matrices $\sigma^{(i)}_{x,y,z}$ for the charge bases of the $i$-th qubit is:
\begin{equation}\label{eq:h2q}
  \begin{aligned}
  H_{2q} &= \sum_{i =1}^{2}\frac{1}{2}(\epsilon_{0}+\epsilon_{i}\sigma_{z}^{(i)}+\Delta_{i}\sigma_{x}^{(i)})+ \\ 
  &\frac{1}{2}(J_{1}I^{(1)}\otimes I^{(2)}+J_{2}\sigma_{z}^{(1)}\otimes\sigma_{z}^{(2)})
\end{aligned},
\end{equation}

where $\epsilon_{0} = 9.4 \cdot 10^{-4} eV$ is the kinetic energy of the electron. $\epsilon_{i}\equiv E_{\text{R},i}-E_{\text{L},i}$ ($i=1,2$) is the detuning of the $i$th qubit, where $E_{\text{R},i}$ and $E_{\text{L},i}$ are the eigenenergies of the state $\ket{R}$ and $\ket{L}$ with a detuning. $\Delta_{i}\equiv E (\Phi_U)_{i} - E(\Phi_L)_{i}$ is the tunneling coupling energy of the $i$-th qubit, where $E_{L,i}$ and $E_{U,i}$ are the eigenenergies of the bonding and anti-bonding like orbits when $\epsilon_i=0$. In this paper, we use two identical TI qubits and thus assume $\Delta_{1}=\Delta_{2}$. $J_i$ are the inter-dot coupling energies between the two TI thin films in the two-qubit bases and $J_i$ can be calculated from the Coulomb interaction between the charge states ${\ket{LL}, \ket{LR}, \ket{RL}, \ket{RR}}$. 

\subsubsection{Device parameters and charge bases} 
The parameters can be calculated from the Hamiltonian $H$ Eq.\ref{eq:cH} . The single qubit parameters $\epsilon_{i}$, $\Delta_{i}$ can be obtained by rewriting the Hamiltonian Eq.\ref{eq:cH} in the basis of the charge states ${\ket{L,\uparrow},\ket{R,\uparrow},\ket{L,\downarrow},\ket{R,\downarrow}}$, where $\ket{L(R),\uparrow(\downarrow)}$ are the single electron states with the charge being localised on the LHS (RHS) , with up (down) spin. The effective Hamiltonian $H_{eff}$ in this basis is:
\begin{equation}
   \begin{aligned}
       &H_{eff} = \int_{-L/2}^{L/2} dz\\&[\ket{L\uparrow},\ket{R\uparrow},\ket{L\downarrow},\ket{R\downarrow}]^{+}H[\ket{L\uparrow},\ket{R\uparrow},\ket{L\downarrow},\ket{R\downarrow}]
\end{aligned}. 
\end{equation}

 Writing $H_{eff}$ using the Pauli matrices, we have:
 \begin{equation}\label{eq:sq}
    H_{eff} = (\epsilon_{0}+\Delta\sigma_x)\otimes I +I\otimes(\epsilon_{0}+\Delta\sigma_x).
 \end{equation}
 
With the external electric field $E(\mathbf{z})= E_0\vec{z}$ and $H_{E} = E(\mathbf{z})I_{4\cross4}$, the effective Hamiltonian $H_{eff,i}'$ of qubit $i$ is:
 \begin{equation}
    H_{eff,i}' = (\epsilon_{i}\sigma_{z}+\Delta\sigma_x)\otimes I +I\otimes(\epsilon_{i}\sigma_{z}+\Delta\sigma_x).
 \end{equation}
 
The detuning $\epsilon_{i}$ is the parameter to be adjusted experimentally to control the two-qubit interaction. 
The Coulomb interaction between the two qubits in the bases of the two-electron charge states is obtained from $U_c$, where:
\begin{equation}
   \begin{aligned}
     &U_{c} = \\
   &\int_{-L/2}^{L/2} dz_{1}\int_{-L/2}^{L/2} dz_{2}\\
   &[\ket{LL},\ket{LR},\ket{RL},\ket{RR}]^{+}U[\ket{LL},
   \ket{LR},\ket{RL},\ket{RR}],
   \end{aligned}
   \end{equation}
   
  where:
  \begin{equation}
 U =\frac{e^2}{4\pi\epsilon_{vac}\epsilon_{r}\sqrt{lx^2+(z_{1}-z_{2})^2}}.
   \end{equation}
  
\begin{figure}[htbp]
    \centering
      \includegraphics[width=0.45\textwidth]{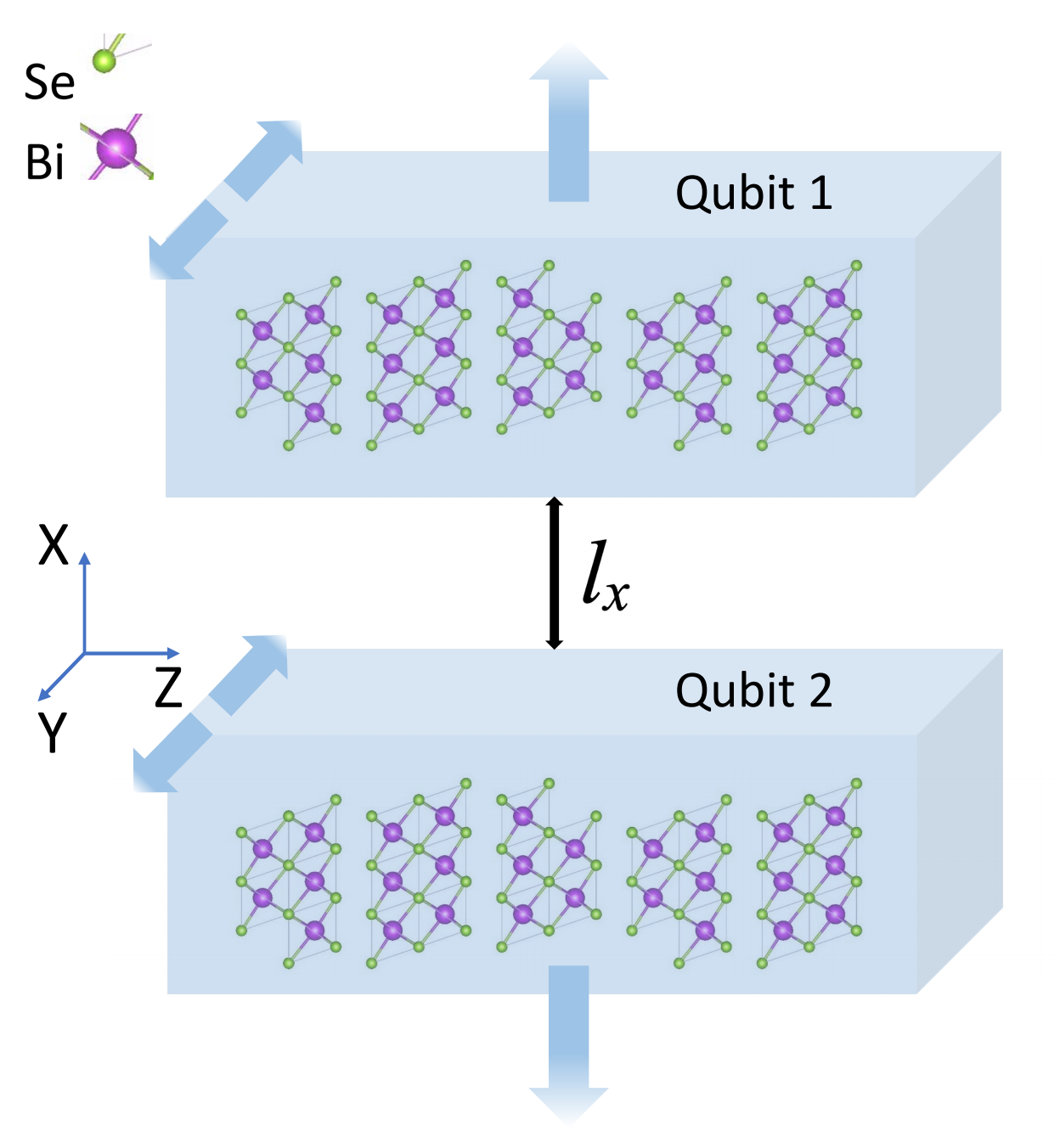} 
        
     \caption{ Setup of the two TI thin films supporting qubits 1 and 2 aligned in parallel with a separation $l_x$, which is finite in the $z$ direction. }
        \label{fig:setup}
\end{figure}

$\epsilon_{vac}$ is the vacuum permittivity, $\epsilon_{r}$ is the dielectric constant of the material, and $l_x$ is the separation between the two TI quantum dots in the $x$-direction (Fig. \ref{fig:setup}).

\subsubsection{The two-qubit operation} 
\begin{figure}[htbp]
     \begin{subfigure}[b]{0.45\textwidth}
    \caption{}
         \includegraphics[width=1.1\textwidth]{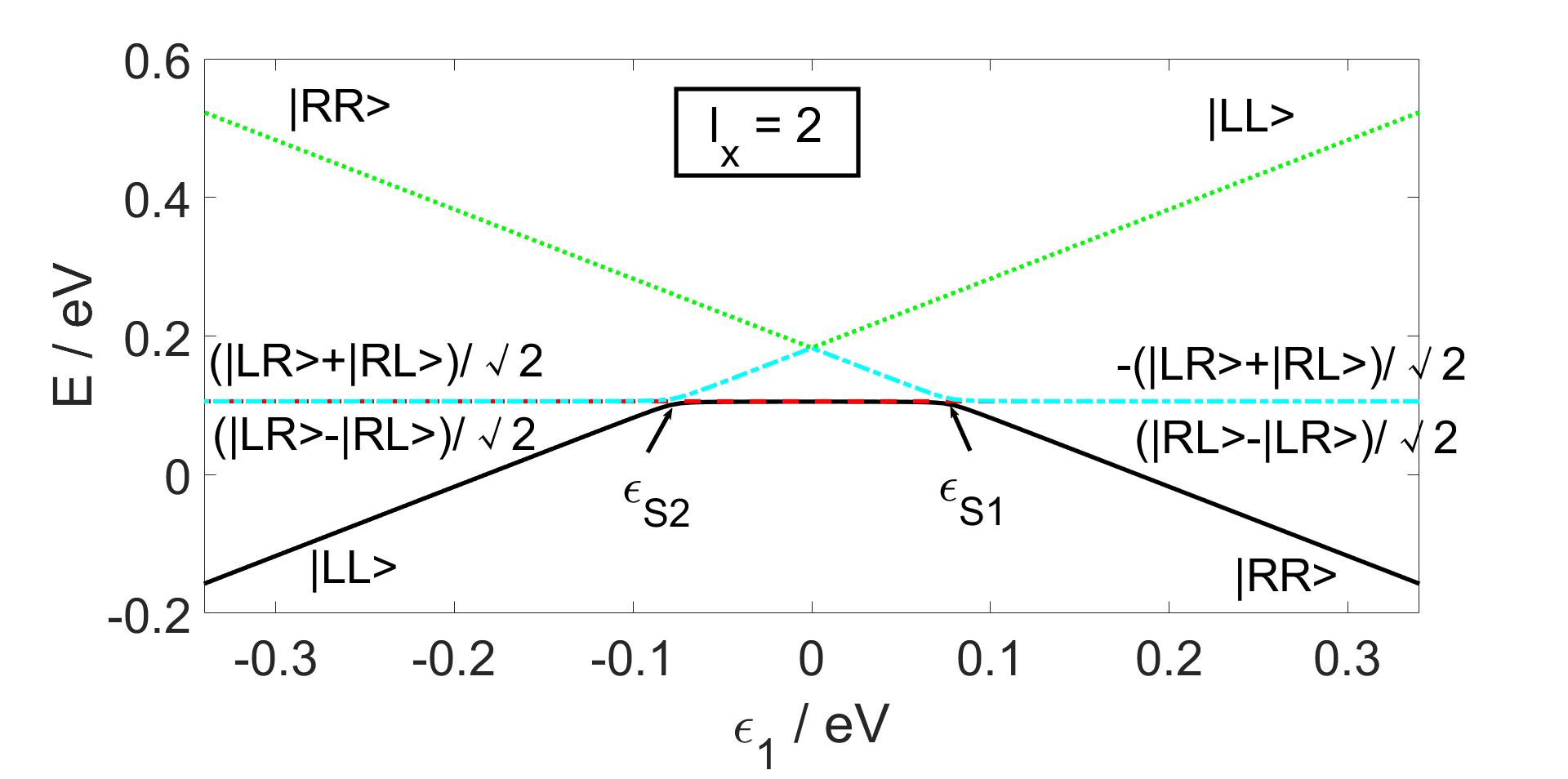}
         \label{fig:lx=2}
     \end{subfigure}
     \begin{subfigure}[b]{0.45\textwidth}
   \caption{}
         \includegraphics[width=1.1\textwidth]{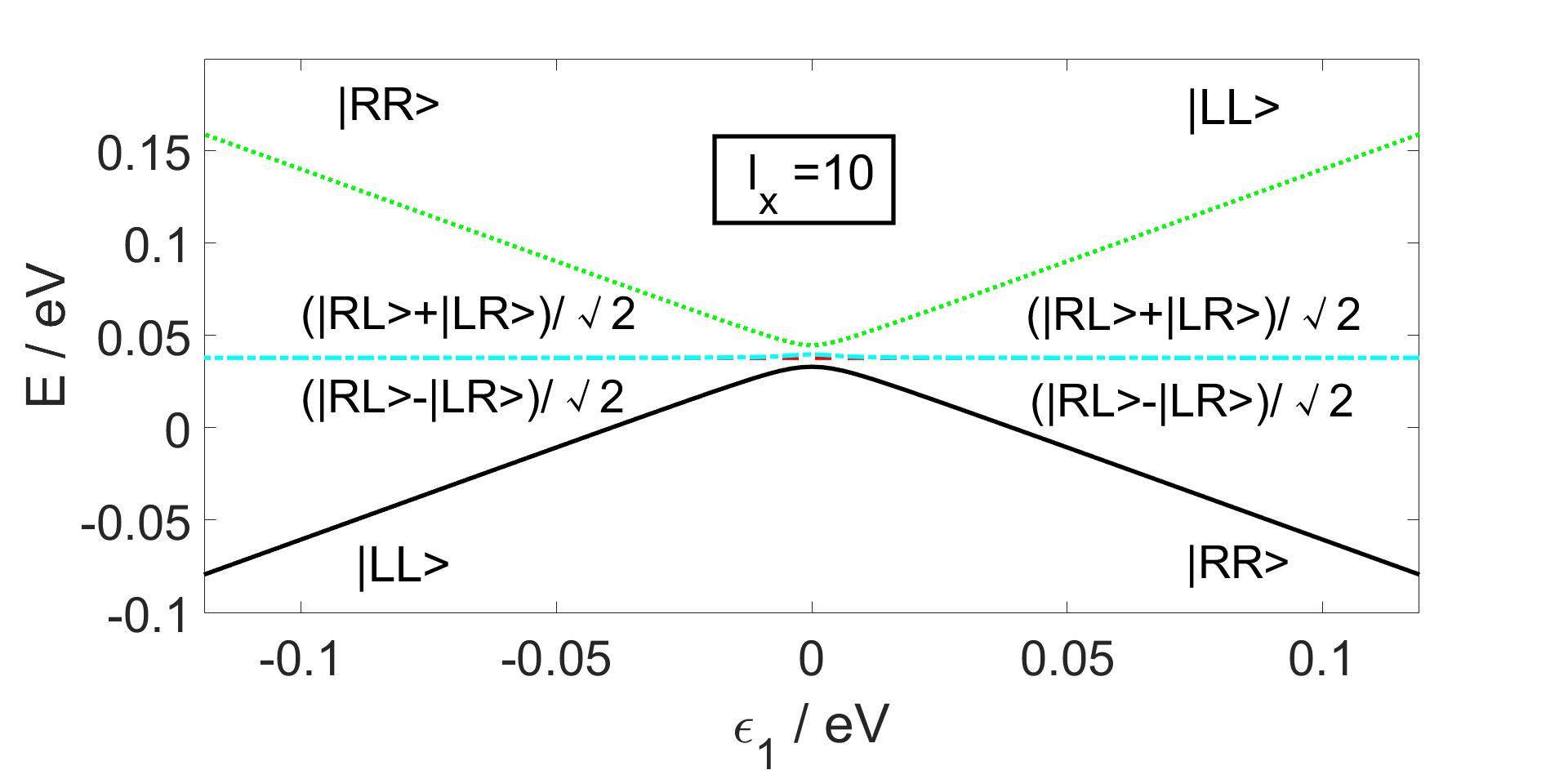}
       
         \label{fig:lx=10}
     \end{subfigure}
     \begin{subfigure}[b]{0.45\textwidth}
   \caption{}
         \includegraphics[width=1.1\textwidth]{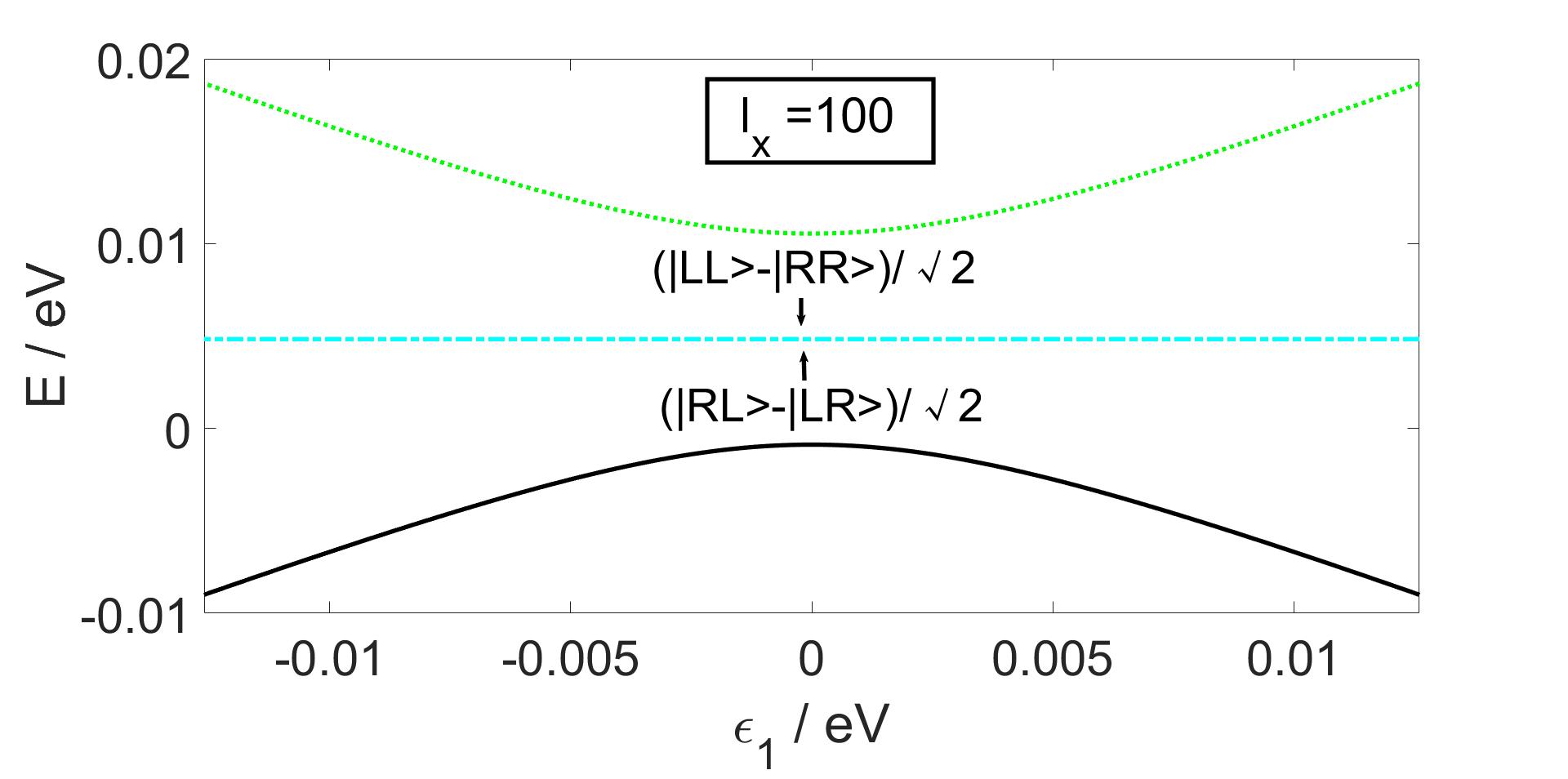}
         \label{fig:lx=100}
     \end{subfigure}
                \caption{Bands of the Hamiltonian Eq.\ref{eq:h2q} along the line $\epsilon_1=\epsilon_2$ with various $l_x$. Band 1 is a black solid line, band 2 a red dashed line, band 3 a cyan dash-dotted line, and band 4 a green dotted line. \textbf{a)} $l_x = \SI{2}{nm}$, $\Delta/J=0.05$. \textbf{b)} $l_x = \SI{10}{nm}$, $\Delta/J=0.15$. \textbf{c)} $l_x = \SI{100}{nm}$, $\Delta/J=1.5$.}
        \label{fig:swap bands}
\end{figure}

The eigenenergies of the two-qubit Hamiltonian $H_{2q}$ vary with $\epsilon_1$ and $\epsilon_2$. The anti-crossings of the energy bands of the states occur when they intersect, which can be seen in Fig. \ref{fig:lx=2}). When $\Delta\gg J$ (where $J = \frac{|J_1-J_2|}{2}$), the onsite interaction $\Delta$ dominates over the Coulomb interaction $J$ and the two-qubit interaction disappears (Fig. \ref{fig:lx=100}). $J \varpropto U_c$, therefore, the distance $l_x$ should not be too large. One way to decrease the ratio of $\Delta/J$ is to increase the thickness $l_z$ of a TI thin film, since $\Delta\equiv E (\Phi_U) - E(\Phi_L)\varpropto 1 / l_z$. 

The time evolution of $H_{2q}$ is written as: $\Omega(t)=e^{-iH_{2q}t/\hbar}$. We can write $H_{2q}$ in its eigenenergy basis using the transformation matrix $A$. Then $H'_{2q}=A^{\dagger}H_{2q}A$ and the time evolution operator $\Omega'(t)$ in the eigenenergy basis of $H_{2q}$ is:
\begin{align}\label{eq:time}
  \Omega'(t) = \begin{pmatrix}
e^{-i\lambda_{1}t/\hbar} & 0 & 0 & 0\\
0 & e^{-i\lambda_{2}t/\hbar} & 0 & 0\\
0& 0 & e^{-i\lambda_{3}t/\hbar} & 0\\
0 & 0 & 0 & e^{-i\lambda_{4}t/\hbar}
\end{pmatrix}, 
\end{align}
where $\lambda_1$, $\lambda_2$, $\lambda_3$ and $\lambda_4$ are the eigenvalues of $H_{2q}$. 

\begin{figure}
     \centering
     \begin{subfigure}[b]{0.5\textwidth}
         \centering
         \caption{}
         \includegraphics[width=\textwidth]{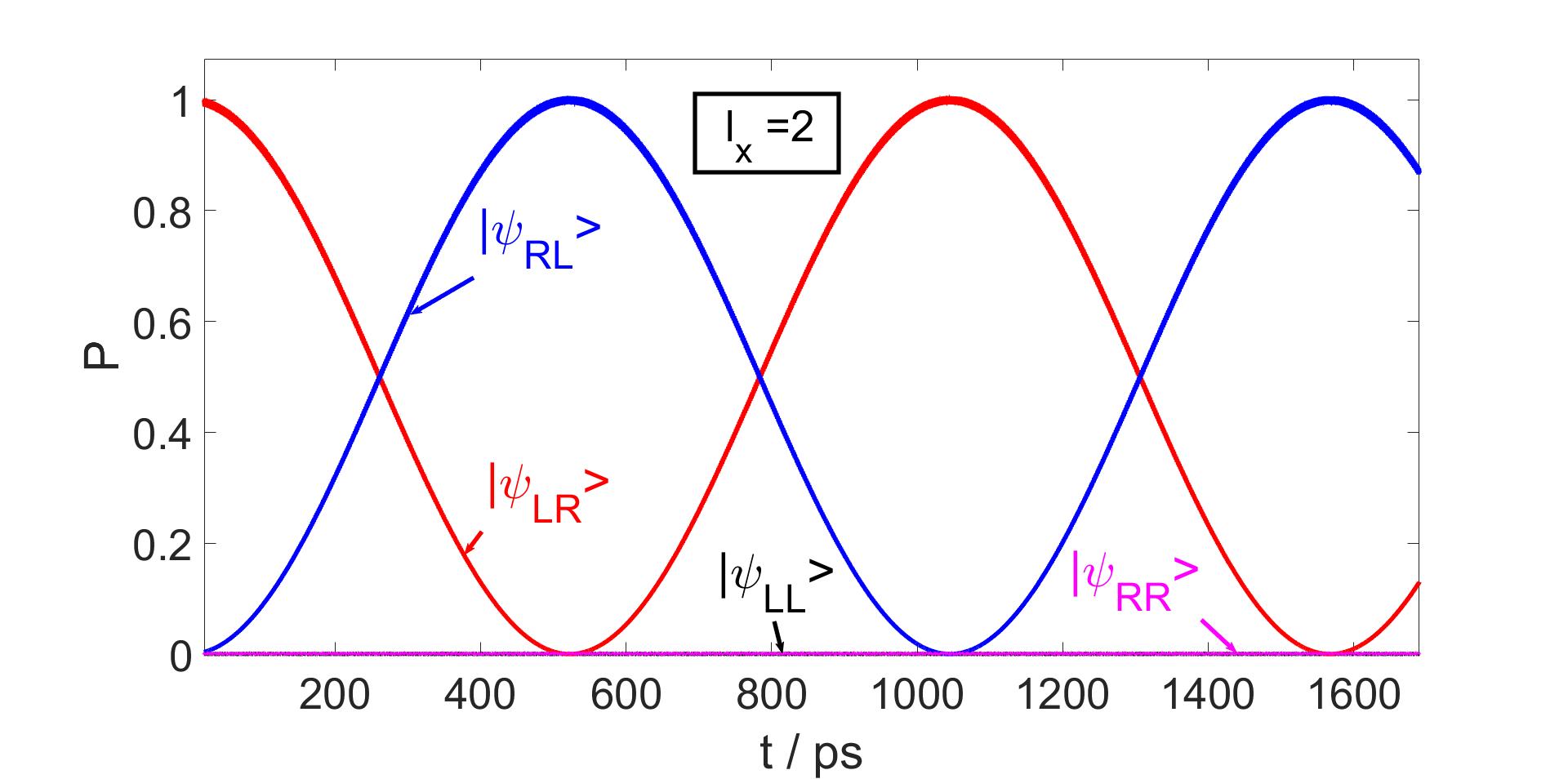}
         \label{fig:teswap}
     \end{subfigure}
     \hfill
     \begin{subfigure}[b]{0.5\textwidth}
         \centering
         \caption{}
         \includegraphics[width=\textwidth]{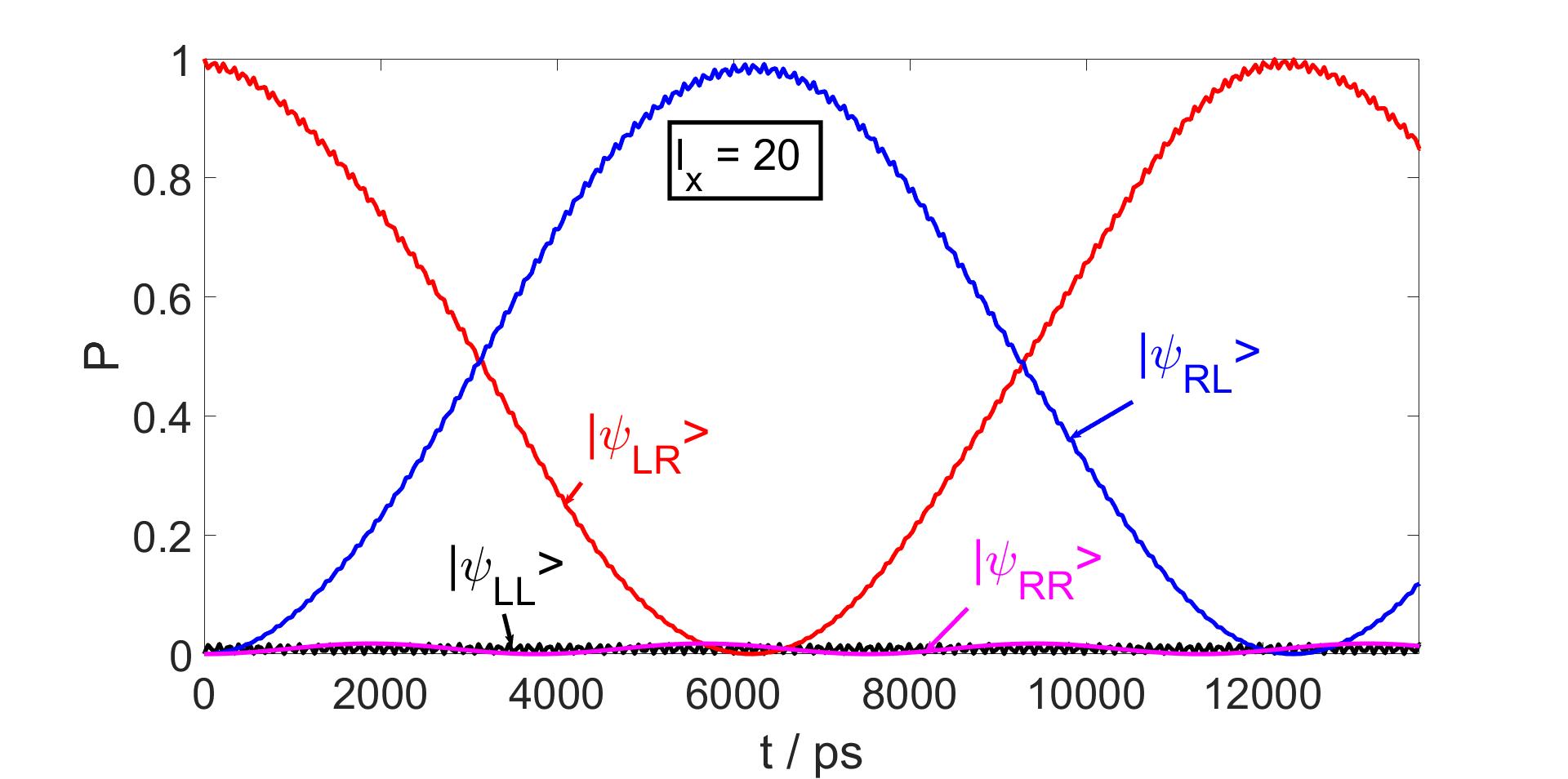}
         \label{fig:tecq}
     \end{subfigure}
     \caption{Time evolution of the state $\ket{LR}$ at various $l_x$. \textbf{a)} Time evolution of the state $\ket{LR}$ at $l_x=\SI{2}{nm}$. A SWAP gate occurs with a period of $\SI{521}{ps}$. \textbf{b)} Time evolution of the state $\ket{LR}$ at $l_x=\SI{20}{nm}$. A SWAP gate occurs with a period of $\SI{6266}{ps}$.}
\end{figure}

SWAP gates exist periodically over a large range of $0 < l_x < \thicksim \SI{80}{nm}$ along the line $\epsilon_{1} = \epsilon_{2}$ (Fig. \ref{fig:teswap}). The period of a SWAP gate increases when $l_x$ increases ( Fig. \ref{fig:tecq}). The unwanted small-amplitude fast oscillations are caused by the off-resonant first order tunneling \cite{fujisawa2011multiple} and can be reduced by increasing $\epsilon_i$ or reducing $l_x$. At $|\epsilon_i| \gg J$, bands 2 and 3 are degenerate. The degeneracy is lifted at the regime around $\epsilon_i=0$ via the intersection of bands 1 and 4 if the Coulomb interaction between the two qubits is strong compared to the onsite tunneling (i.e. if $\Delta/J$ is small). No SWAP gate could be found if there is no degeneracy splitting. In Fig. \ref{fig:lx=100} where $l_x =  \SI{100}{nm}$, $\Delta/J = 1.5$, it can be seen that bands 1 and 4 detach from the middle bands at the area of $\epsilon_1$ around zero and there is no valid SWAP gate. In Fig. \ref{fig:lx=10} where $l_x=\SI{10}{nm}$, $\Delta/J=0.15$, there is a small regime of resonance of bands 1 and 4 with the middle bands which splits the degeneracy of bands 2 and 3. SWAP gates are observed in the regime of $\epsilon_1$ away from zero, where bands 2 and 3 are degenerate. In Fig. \ref{fig:lx=2} where $l_x=\SI{2}{nm}$, $\Delta/J=0.05$, bands 1 and 2 are degenerate in the regime $\epsilon_{S2}<\epsilon_1 <\epsilon_{S1}$ ($\epsilon_{S1}$ and $\epsilon_{S2}$ are the anti-crossing points when band 1 (black) intersects band 3 (cyan) in Fig. \ref{fig:lx=2}, which should not be confused with the detunings $\epsilon_i$. $\epsilon_{S1}=-\epsilon_{S2}$ since $\ket{L}$ and $\ket{R}$ are symmetric about the middle of the TI thin film along the $z$ direction). SWAP gates are observed in the regime $\epsilon_1>\epsilon_{S1}$ and $\epsilon_1<\epsilon_{S2}$. It should be noted that SWAP gates exist in the regime where bands 2 and 3 are degenerate. At half the period of a SWAP gate ($T_{\text{SWAP}}$), we obtain a $\sqrt{\text{SWAP}}$ gate, which produces two entangled Bell states Eq.\ref{eq: teswap1} and Eq.\ref{eq: teswap2} as desired (Fig. \ref{fig:teswap}).

\begin{align}
&\ket{LL}\xrightarrow[]{\Omega(T_{\text{SWAP}}/2)}\ket{LL},\\
 &\ket{LR}\xrightarrow[]{\Omega(T_{\text{SWAP}}/2)}\frac{1}{\sqrt{2}}(\ket{LR}+\ket{RL}),\label{eq: teswap1}\\
 &\ket{RL}\xrightarrow[]{\Omega(T_{\text{SWAP}}/2)}\frac{1}{\sqrt{2}}(\ket{LR}-\ket{RL}),\label{eq: teswap2}\\
 &\ket{RR}\xrightarrow[]{\Omega(T_{\text{SWAP}}/2)}\ket{RR}.
  \end{align}

\subsubsection{Floquet charge qubit}
\begin{figure*}[htbp]
     \begin{subfigure}[b]{0.45\textwidth}
         \caption{}
         \includegraphics[width=1.1\linewidth]{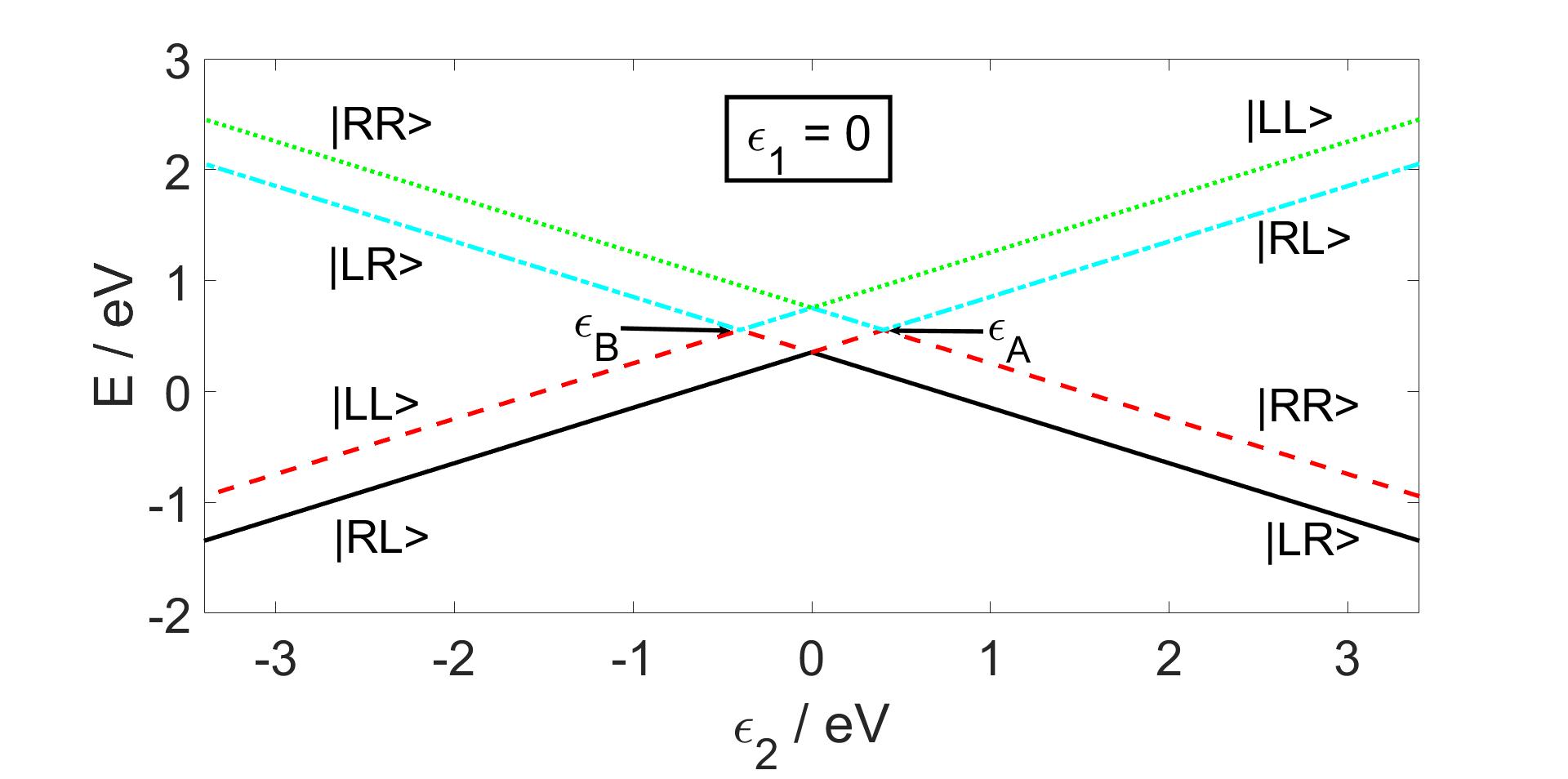}   
         \label{fig:e1zero}
     \end{subfigure}
     \begin{subfigure}[b]{0.45\textwidth}
 \caption{}
         \includegraphics[width=1.1\textwidth]{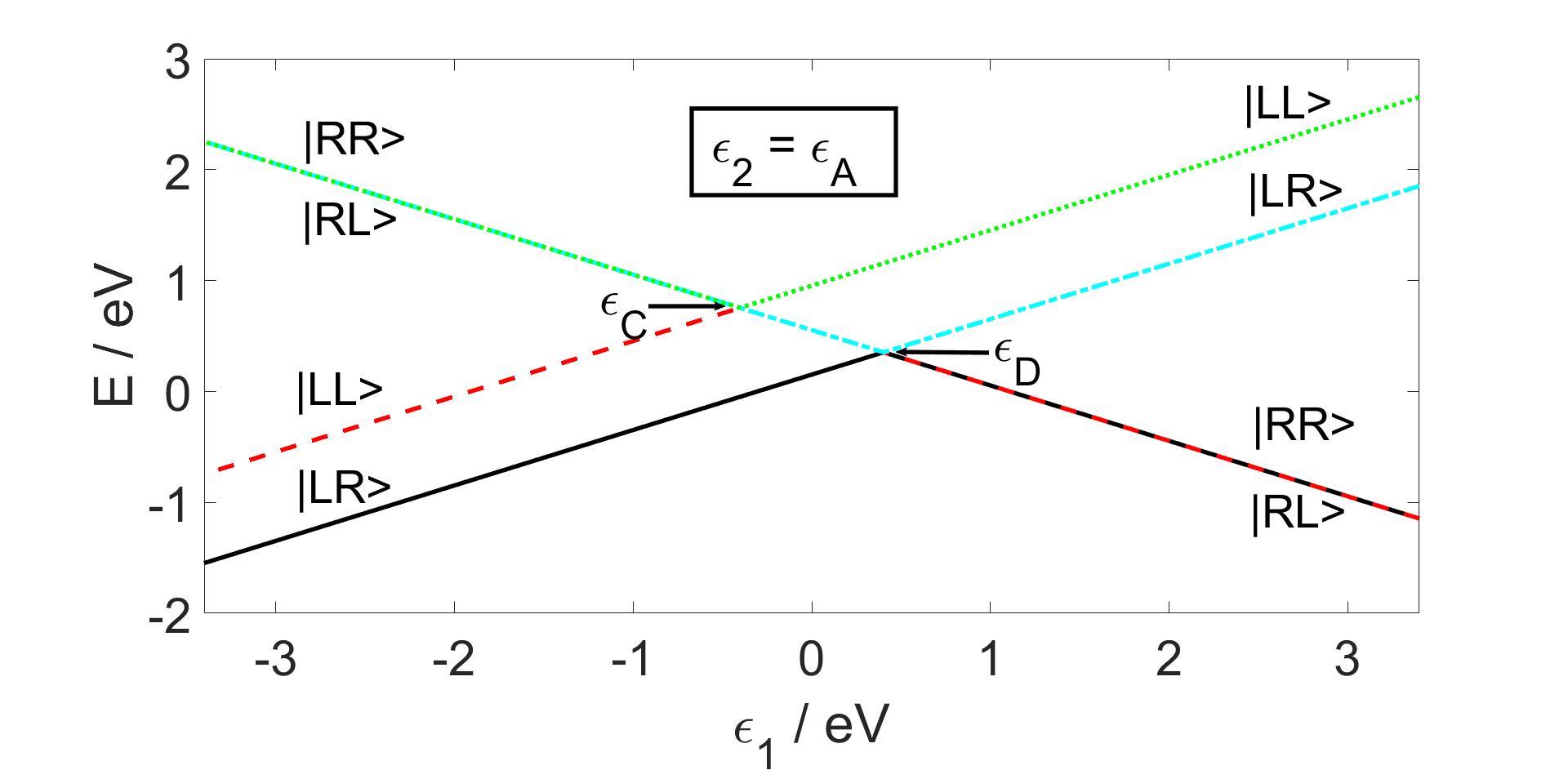}     
         \label{fig:e2a}
     \end{subfigure}
     \begin{subfigure}[b]{0.45\textwidth}
         \caption{}
         \includegraphics[width=1.1\textwidth]{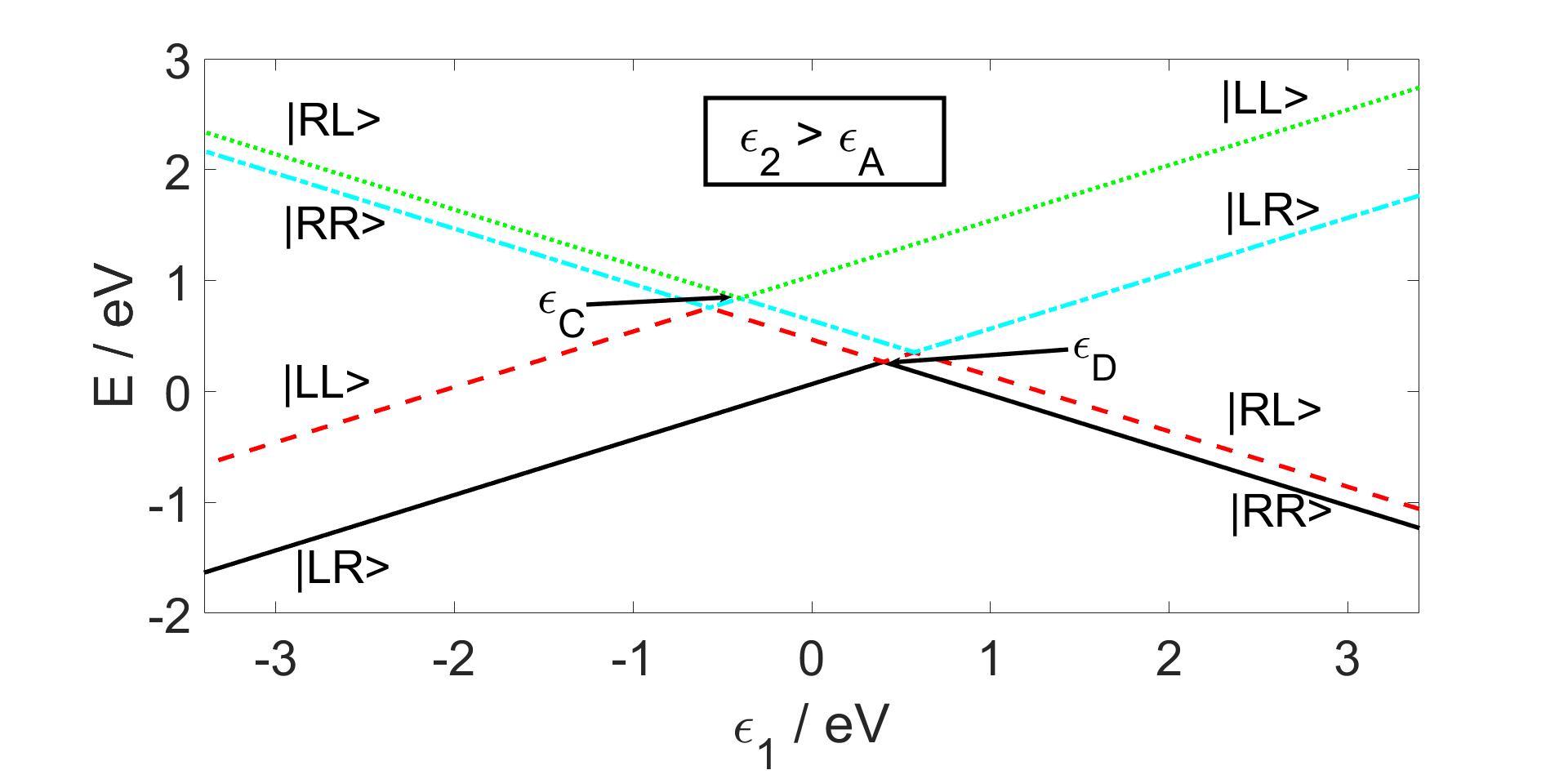}   
         \label{fig:e2la}
     \end{subfigure}
      \begin{subfigure}{0.45\textwidth}
       \caption{}
         \includegraphics[width=1.1\textwidth]{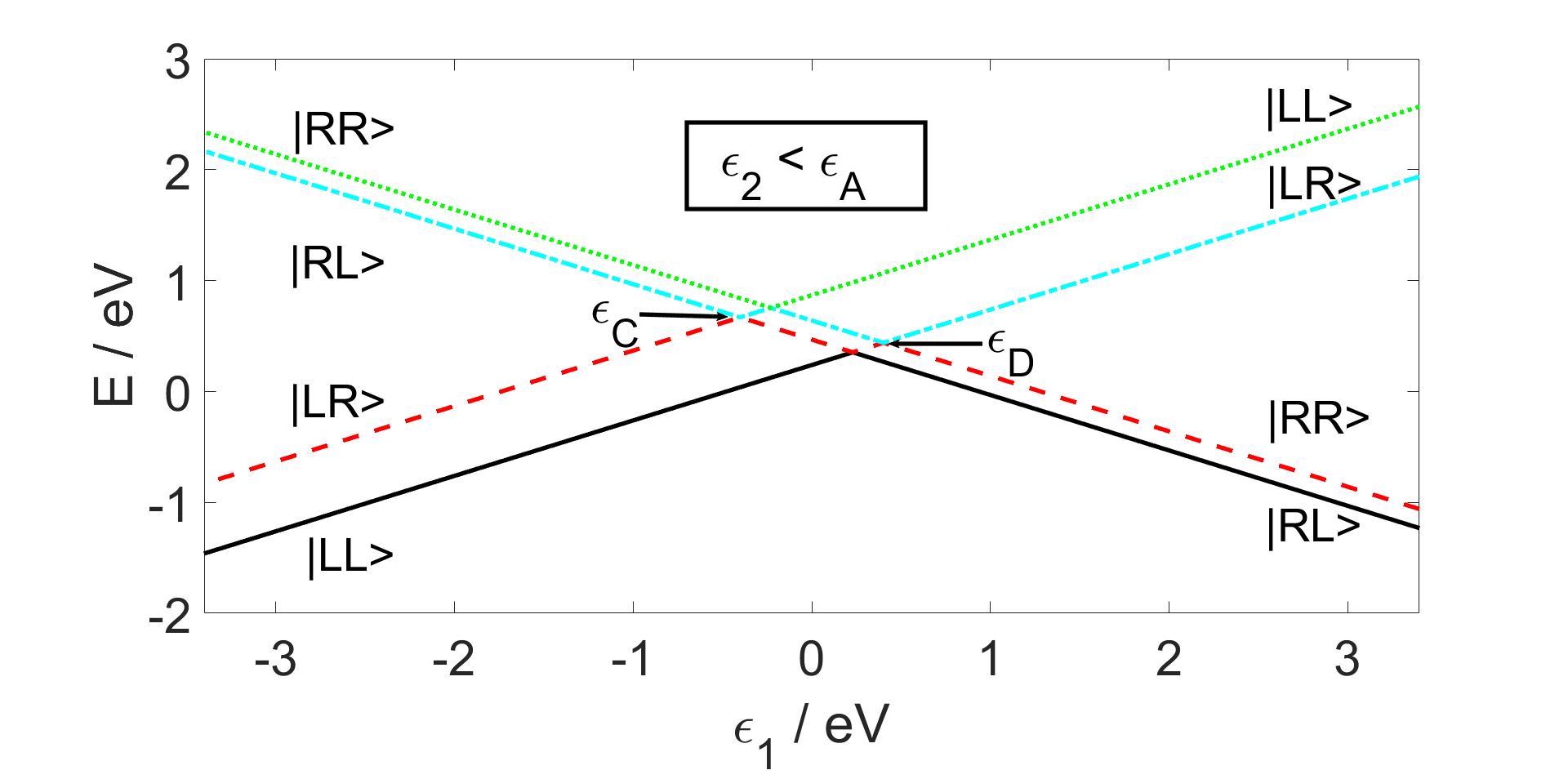}
         \label{fig:e2sa}
     \end{subfigure}
        \caption{Bands of the Hamiltonian Eq. \ref{eq:hf2q}  at $l_x = \SI{0.5}{nm}$ vs $\epsilon_i$. The bands remain in the Floquet states over the range of $\epsilon_i$. Band 1 is in black solid line, band 2 in red dashed line, band 3 in cyan dash-dotted line, and band 4 in green dotted line. \textbf{a)} Along the line $\epsilon_{1} = 0$. \textbf{b)} Along the line $\epsilon_{2} = \SI{0.40}{eV}=\epsilon_{A}$. \textbf{c)} Along the line $\epsilon_{2} = \SI{0.57}{eV} > \epsilon_{A}$.}
        \label{fig:bandsflo}
\end{figure*}

The two-qubit Hamiltonian is slightly different from the bare TI system Eq. \ref{eq:h2q}. The Floquet solutions ${\ket{\Phi_{\text{FL}},\uparrow},\ket{\Phi_{\text{FR}},\uparrow},\ket{\Phi_{\text{FL}},\downarrow},\ket{\Phi_{\text{FR}},\downarrow}}$ are used as the charge basis, where $\ket{\Phi_{\text{FL}}(\Phi_{\text{FR}}),\uparrow(\downarrow)}$ are the single Floquet states with the charge being localised on the LHS (RHS) , with up (down) spin, and they are the eigenstates of Eq. \ref{eq:cH} with an oscillating periodic electric field $E_{0}\cos{wt}$ (the same field used in Sec. \ref{sec:flqu}). Instead of Eq. \ref{eq:sq} , we have:
 \begin{equation}
    H_{Feff} = E_{1}I\otimes I-E_{2}I\otimes\sigma_{z},  
 \end{equation}
 where 
$E_{1}=\frac{E(\Phi_{\text{FL},\uparrow})+E(\Phi_{\text{FL},\downarrow})}{2}$, $E_{2}=\frac{E(\Phi_{\text{FL}.\uparrow})-E(\Phi_{\text{FL},\downarrow})}{2}$.  $E(\Phi_{\text{FL},\uparrow})$ and $E(\Phi_{\text{FL},\downarrow})$ are the Floquet energies of the Floquet states. 
 
 The two-qubit Hamiltonian in the Floquet charge qubit basis is:
 \begin{equation}\label{eq:hf2q}
  \begin{aligned}
  H_{f2q} &= \sum_{i =1}^{2}\frac{1}{2}(\epsilon_{i}\sigma_{z}^{(i)}+E(\Phi_{FL,\uparrow})I^{(i)} ) \\
  &+\frac{1}{2}(J_{1}I^{(1)}\otimes I^{(2)}+J_{2}\sigma_{z}^{(1)}\otimes\sigma_{z}^{(2)}).
  \end{aligned}
\end{equation}
 
 The detuning $\epsilon_{i}$ is the parameter to be adjusted experimentally to control the Floquet two-qubit interactions.
 
\subsubsection{The two-qubit operation} 
\begin{figure}[htbp]
     \centering
     \begin{subfigure}[b]{0.5\textwidth}
         \centering
          \caption{}
         \includegraphics[width=\textwidth]{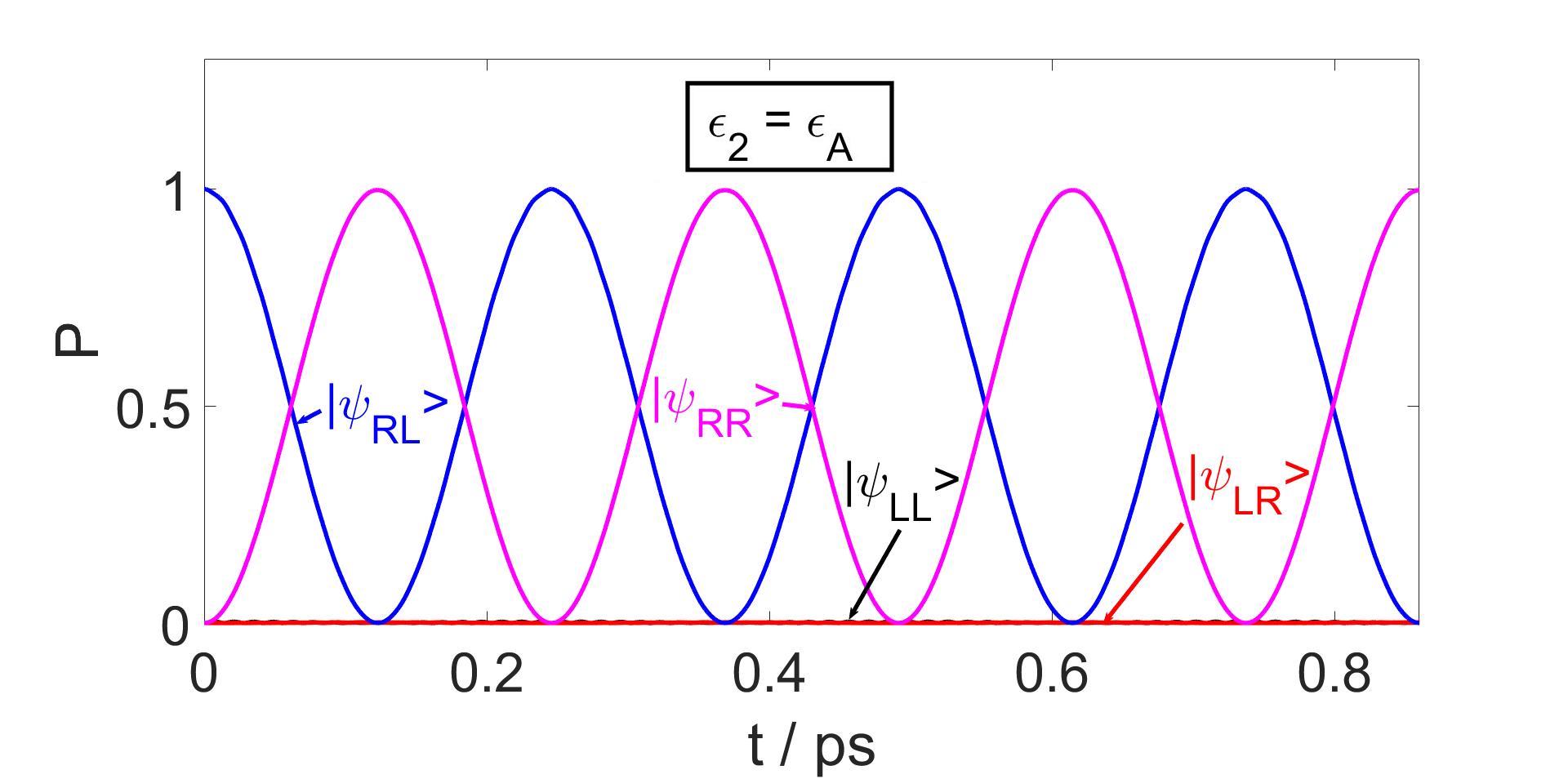}    
         \label{fig:tee2a}
     \end{subfigure}
     \hfill
     \begin{subfigure}[b]{0.5\textwidth}
       \centering
        \caption{}
         \includegraphics[width=\textwidth]{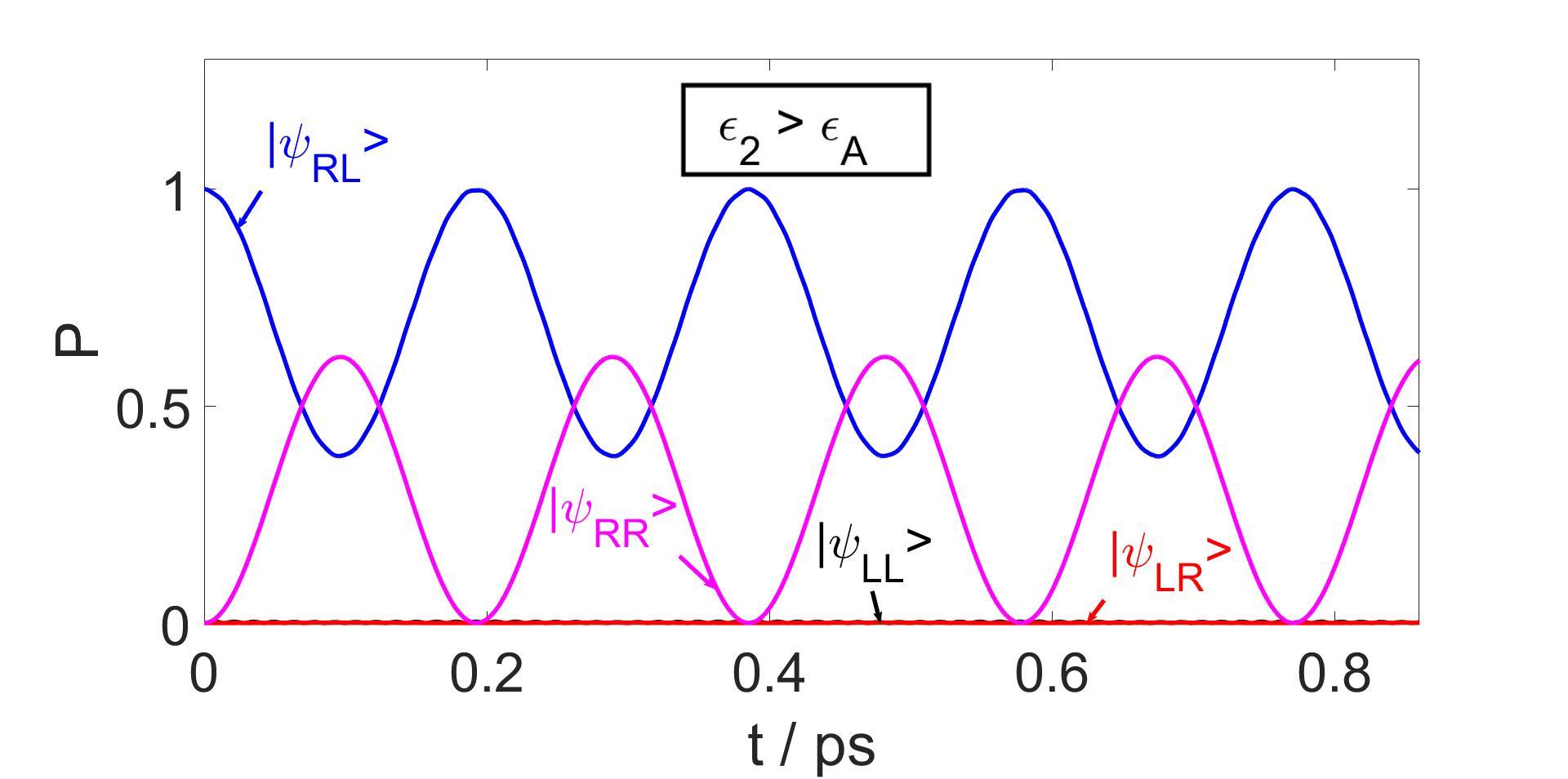}         
         \label{fig:tee2la}
     \end{subfigure}
     \hfill
     \begin{subfigure}[b]{0.5\textwidth}
         \centering
           \caption{}
         \includegraphics[width=\textwidth]{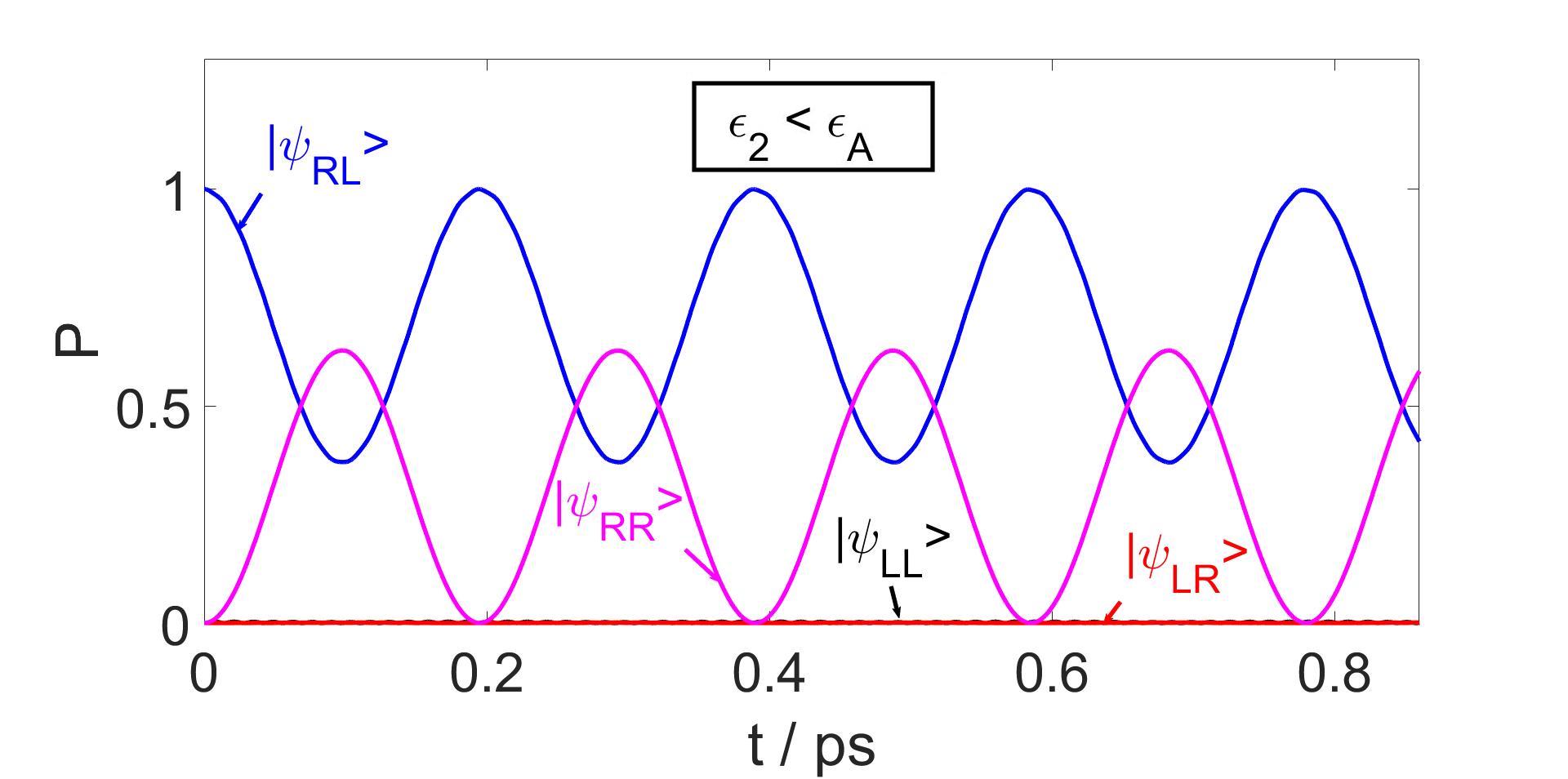}  
         \label{fig:tee2sa}
     \end{subfigure}
     \caption{Time evolution of the state $\ket{RL}$ at various $\epsilon_2$. \textbf{a)} Time evolution of the state $\ket{RL}$ at $\epsilon_2=\epsilon(A) = \SI{0.40}{eV}$. CROT operations have a period of $\SI{0.25}{ps}$. A CNOT gate is observed at $\SI{0.12}{ps}$. \textbf{b)} Time evolution of the state $\ket{RL}$ at $\epsilon_2 = \SI{0.57}{eV} > \epsilon(A)$. CROT operations have a period of $\SI{0.19}{ps}$. \textbf{c)} Time evolution of the state $\ket{RL}$ at $\epsilon_2 = \SI{0.23}{eV} < \epsilon(A)$. CROT operations have a period of $\SI{0.19}{ps}$.}
\end{figure}

Controlled rotation (CROT) operations are observed at a small separation $l_x=\SI{0.5}{nm}$ for the same setup using a pair of Floquet charge qubits, when the target qubit has a detuning parameter $\epsilon_i=\epsilon_{A}$ or $\epsilon_{B}$ as shown in Fig. \ref{fig:e1zero}, where $\epsilon_{A}$ and $\epsilon_{B}$ are the crossing points of bands 2 (red) and 3 (cyan) in the plot. The half period of the CROT operation provides a CNOT gate Fig. \ref{fig:tee2a}. The state ($\ket{L}$ or $\ket{R}$) of the control qubit is chosen by the sign of the detuning (thus the direction of the electric field) of the target qubit. From Fig. \ref{fig:e2a}, when $\epsilon_2 = \epsilon_{A}$, we have a CROT operation which rotates the state of qubit 2 only when the state of qubit 1 is $\ket{R}$. When $\epsilon_2 = \epsilon_{B}$, a CROT operation rotates the state of qubit 2 only when the state of qubit 1 is in $\ket{L}$. In Fig. \ref{fig:e2a} where $\epsilon_{2}=\epsilon_{A}$, the relevant bands $\ket{RR}$ and $\ket{RL}$ are degenerate and at maximum resonance in the regime $\epsilon_{1} < \epsilon_{C}$ or $\epsilon_{1} > \epsilon_{D}$, where $\epsilon_{C} = -\SI{0.40}{eV}$ and $\epsilon_{D} = +\SI{0.40}{eV}$ are the crossing points in the plot. If the target detuning $\epsilon_2$  moves away from $\epsilon_{A}$, the relevant states are separated (Fig. \ref{fig:e2la} and Fig. \ref{fig:e2sa} ) and the range of the angles of rotation of the CROT operation decreases (Fig. \ref{fig:tee2la} and Fig. \ref{fig:tee2sa}). The fidelity of the CROT operation increases with increasing $|\epsilon_{1}|$ (the detuning of the control qubit). In this paper, we use two identical TI thin films with the same periodic oscillating electric field applied on each. Therefore, the two qubits are identical, and $\epsilon_{A} = \epsilon_{D}$, $\epsilon_{B} = \epsilon_{C}$. Because $\ket{L}$ and $\ket{R}$ are symmetric about the middle of the TI thin film along the $z$ direction, $\epsilon_{B}= -\epsilon_{A}$ and $\epsilon_{C}= -\epsilon_{D}$. 
\section{CONCLUDING REMARKS AND OUTLOOKS}
In this paper, we investigated a charge qubit based on a new material - a 3D topological insulator. We have proposed a complete implementation scheme for initializing, operating single and two-qubit quantum gates, reading out the qubits. Moreover, we studied a Floquet-engineered TI qubit and found that it can be initialized, operated on using single and two-qubit quantum gates, and readout as well. We conclude that it is possible to use an ultra-thin TI system for universal quantum computing. In the article, we consider a theoretical TI device without the effect of temperature and the electron-electron interactions. In a realistic device, the relaxations and the electron electron interaction may cause the decoherence of the qubit state. These effects in an ultra-thin TI system are under active studies \cite{PhysRevMaterials.6.044203,ZHAO2022506,lu2014weak}. The relaxation and the electron-electron interaction processes are found to be temperature dependent, which indicates possible phonon-assisted process \cite{PhysRevMaterials.6.044203,ZHAO2022506}. Therefore, the operating temperature should be concerned in a realistic TI devices. In an ultra-thin TI system, the berry phase is reduced from $\pi$ gradually with the decrease of the thickness \cite{PhysRevB.95.235429}. However, the anti weak-localization still exists, say in a $\SI{5}{QL}$ TI system. This implies that the system is still robust against scattering of impurities to some extend, which could be beneficial to its decoherence time. Also, the TI devices are in nanometers, which is advantageous in terms of fabricating compact and large-scale quantum circuits. With these benefits, we think TIs would be a promising candidate for the future of fault-tolerant quantum computing.   
\section*{Acknowledgements}
 This work was supported by the China Scholarship Council. The authors would like to thank Dr Tianwei Wang for their help in creating figures 1 and 5. The authors declare no conflict of interest.

\bibliographystyle{unsrt} 
\cleardoublepage
\bibliography{References/references} 

\end{document}